\documentclass[twoside,12pt]{article}

\usepackage{epsfig}
\usepackage{rotating}
\usepackage{axodraw}
\usepackage{amsmath}
\usepackage{amssymb}
\usepackage[dvips]{color}

\usepackage[normalem]{ulem}
\usepackage{xcolor}
\usepackage{physics}
\usepackage[symbol]{footmisc}
\renewcommand{\thefootnote}{\fnsymbol{footnote}}

\newcommand{\tgb}{{\rm tg}\beta}
\newcommand{\tga}{{\rm tg}\alpha}
\newcommand{\Quark}[4]{\ArrowLine(#1,#2)(#3,#4)}
\newcommand{\Gluino}[5]{\Gluon(#1,#2)(#3,#4){3}{#5}\Line(#1,#2)(#3,#4)}
\newcommand{\CrossedCircle}[6]{\BCirc(#1,#2){5}\Line(#3,#6)(#5,#4)\Line(#5,#6)(#3,#4)}
\newcommand{\tR}{t_R}
\newcommand{\tL}{t_L}
\newcommand{\ts}{\widetilde{t}}

\newcommand{\tsL}{\ts_L}
\newcommand{\tsR}{\ts_R}
\newcommand{\gl}{\widetilde{g}}
\newcommand{\gluino}{\widetilde{g}}
\newcommand{\st}{\tilde{t}}
\newcommand{\sbottom}{\tilde{b}}
\newcommand{\lt}{\lambda_t}
\newcommand{\stgb}{{\rm tg}^2\beta}
\newcommand{\sia}{\sin\alpha}
\newcommand{\coa}{\cos\alpha}
\newcommand{\sib}{\sin\beta}
\newcommand{\cob}{\cos\beta}
\newcommand{\sgl}{\tilde{g}}

\newcommand{\uus}{\widetilde{U}}
\newcommand{\dds}{\widetilde{D}}

\begin{document}

\renewcommand{\thefootnote}{\fnsymbol{footnote}}

\renewcommand{\thefootnote}{\arabic{footnote}}

\setcounter{footnote}{0}

\vspace*{-1.0cm}

\begin{flushright}
KA--TP--12--2023 \\
PSI--PR--23--18 \\
CERN--TH--2023--112
\end{flushright}

\begin{center}
{\large \sc Charged Higgs-Boson Decays into Quarks}
\\[0.5cm]

Jamie Chang$^1$, Fiona Kirk$^{2,3}$, Margarete M\"uhlleitner$^4$ and Michael
Spira$^1$ \\[0.3cm]

{\it $^1$ Paul Scherrer Institut, CH--5232 Villigen PSI, Switzerland} \\
{\it $^2$ Physikalisch-Technische Bundesanstalt, Quantum Frontiers,
          D--38116 Braunschweig, Germany} \\
{\it $^3$ Theoretical Physics Department, CERN, CH--1211 Gen\'eve 23,
          Switzerland} \\
{\it $^4$ Institute for Theoretical Physics, Karlsruhe Institute of
          Technology, D--76128 Karlsruhe, Germany}
\end{center}

\begin{abstract}
\noindent
We consider the full genuine next-to-leading order SUSY--QCD corrections
to the charged Higgs decays into quarks supplemented by the NNLO
corrections to the effective top and bottom Yukawa couplings. The NNLO
corrections to the effective top Yukawa coupling are a new ingredient of
our analysis. We arrive at an approximate NNLO prediction for MSSM
charged Higgs decays after including the N$^4$LO QCD corrections for
large charged Higgs masses. The residual uncertainties are in the
percent range or below, depending on the particular MSSM scenario.
\end{abstract}

\section{Introduction}
The Standard Model (SM) of particle physics has been very successful in
describing scattering and decay processes from the high- to the low-energy
frontiers \cite{pdg}. A cornerstone of this model is the Higgs mechanism
for electroweak symmetry breaking \cite{higgs}. The discovery of the
Higgs boson in 2012 \cite{discovery,couplings} has confirmed this
realization of spontaneous symmetry breaking and
completed the required particle content of the SM \cite{sm}. Although
the SM describes all processes at the high-energy frontier and shows the
proper behaviour of processes in the high-energy limit, it cannot
describe all observations. In particular, the evidence for Dark Matter
and the inability of the SM to generate the baryon asymmetry of the
universe call for physics beyond the SM.

The Higgs sector allows the electroweak force to remain weakly
interacting up to very high-energy scales \cite{smren,unitarity}.  One
of the major lines of research beyond the SM is the formulation of grand
unified theories (GUTs) which are broken down to the low-energy SM at an
energy scale of
the order of $10^{16}$ GeV and can be motivated by the approximate
unification of the gauge couplings of the SM. This requires the SM to be
weakly interacting up to these high-energy scales, which is only
possible if the Higgs mass takes a value in the range between about 130
and 190 GeV \cite{hmassbounds}. This is compatible with the observed
Higgs mass of $(125.09\pm 0.24)$ GeV, if the universe is allowed to be
metastable \cite{metastability}.

However, even if the Higgs mass is in this range, a further problem
arises. If the SM is embedded in a GUT, quadratic divergences in
higher-order corrections to the Higgs mass suggest that, if the Higgs boson
couples directly or indirectly to particles with masses at the GUT
scale, quantum fluctuations tend to raise the Higgs mass to this new
mass scale \cite{hierarchy}. To stabilize the Higgs mass at the
electroweak scale, an extreme fine-tuning of the corresponding mass
counterterm is necessary. This hierarchy problem can be avoided by the
introduction of supersymmetry \cite{susy}, a novel symmetry between the
bosonic and fermionic degrees of freedom of the model. Supersymmetric
models are free of quadratic divergences and solve the hierarchy problem
if the superpartners of all SM particles acquire masses below about
$1-10$ TeV. Supersymmetric GUTs predict a value of the Weinberg angle
that is in striking agreement with the precision measurements at LEP and SLC
\cite{sw2sgut}.  The minimal supersymmetric extension of the SM (MSSM)
\cite{mssm,mssm1} requires the existence of 5 elementary Higgs bosons.
These consist of two neutral {\cal CP}-even states $h,H$, one neutral
{\cal CP}-odd $A$ state, and two charged Higgs states $H^\pm$.  For real MSSM
parameters, the Higgs sector can be described by two input parameters at
leading order (LO): the pseudoscalar mass $M_A$ and $\tgb$. The latter
is the ratio of the two vacuum expectation values involved in the
neutral components of the two Higgs doublets.

An immediate sign of an extended Higgs sector beyond the SM would be the
discovery of a charged Higgs boson. The main search channel for heavy
charged Higgs bosons is its decay into a $t\bar{b}$ final state. This work
addresses charged Higgs-boson decays into heavy quarks. The decay
$H^+\to t\bar b$ provides the dominant charged Higgs decay channel for a
large range of the charged Higgs mass, depending on the MSSM scenario. In
our numerical analysis, we have adopted the $M_h^{125}$ benchmark
scenario of Ref.~\cite{benchmark} as a representative case of the MSSM.
This scenario is not excluded by the experimental searches. The input
parameters defined in the on-shell scheme of Ref.~\cite{benchmark} are
\begin{eqnarray}
\mbox{$M_h^{125}$:} && M_{\tilde Q_3} = 1.5~{\rm TeV},\quad
M_{\tilde \ell_3} = 2~{\rm TeV},\quad M_{\sgl} = 2.5~{\rm TeV},
\nonumber \\
&& M_1 = M_2 = 1~{\rm TeV},\quad A_b = A_\tau = A_t = 2.8~{\rm
TeV} + \mu/\tgb,\quad \mu = 1~{\rm TeV} \, ,
\end{eqnarray}
where $M_{\tilde Q_3}$ ($M_{\tilde \ell_3}$) denotes the (common) left-
and right-handed soft SUSY-breaking mass parameter of the third
generation squarks (sleptons), $M_{\sgl}$ is the gluino mass, $M_1, M_2$
are the soft SUSY-breaking gaugino mass parameters, $\mu$ is the
higgsino mass parameter and $A_b, A_t, A_\tau$ are the soft
SUSY-breaking trilinear coupling parameters of the third generation. We
have determined all squark masses according to the procedure described
in Ref.~\cite{gganlo}. This leads to the following values of the stop
and sbottom masses for two values of $\tgb=10,40$:
\begin{eqnarray}
&& \underline{\tgb=10} \nonumber \\
&& m_{\st_1} = 1340~{\rm GeV}, \quad m_{\st_2} =
1662~{\rm GeV}, \quad m_{\sbottom_1} = 1496~{\rm GeV}, \quad
m_{\sbottom_2} = 1508~{\rm GeV} \nonumber \\
&& \underline{\tgb=40} \nonumber \\
&& m_{\st_1} = 1340~{\rm GeV}, \quad m_{\st_2} =
1662~{\rm GeV}, \quad m_{\sbottom_1} = 1479~{\rm GeV}, \quad
m_{\sbottom_2} = 1525~{\rm GeV}.
\end{eqnarray}
We have used the renormalization-group-improved two-loop corrected
charged Higgs mass and couplings of Ref.~\cite{rgi} in our analysis. The
top pole mass has been chosen as $m_t=172.5$ GeV, the bottom
$\overline{\rm MS}$ mass as $\overline{m}_b(\overline{m}_b) = 4.18$ GeV
and the strong coupling as $\alpha_s(M_Z)=0.118$ in the $\overline{\rm
MS}$ scheme. However, for the bottom mass involved in the derivation of the
sbottom sector, the derived bottom mass has been employed to
avoid large uncancelled ${\rm tg}\beta$-enhanced contributions (see
discussion in Refs.~\cite{sbottom}).

\begin{figure}[htb]
\vspace*{-2.8cm}
\hspace*{-2.0cm}
\epsfig{file=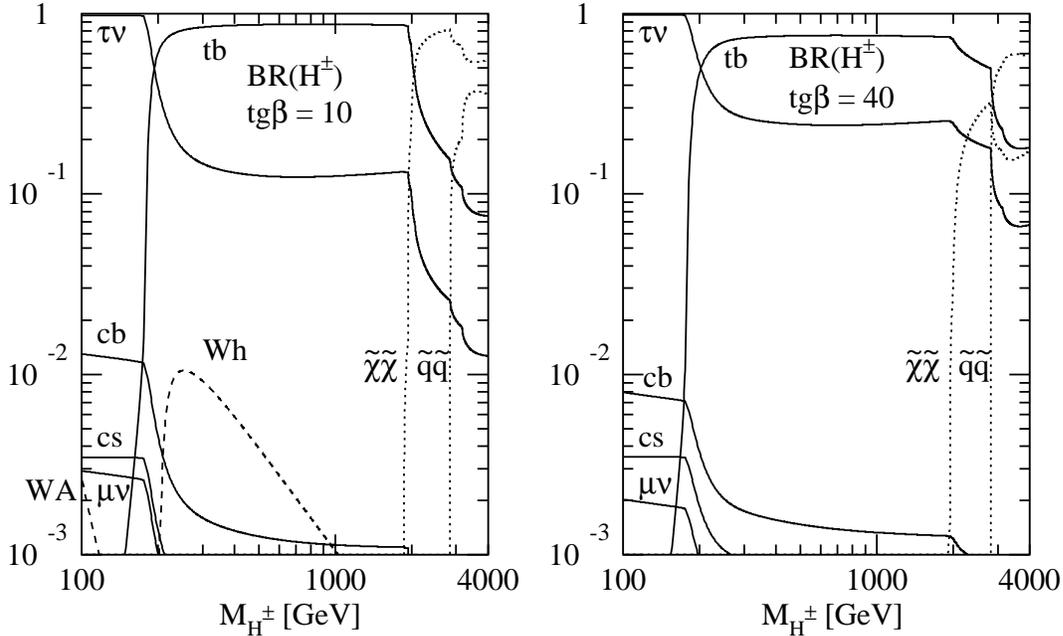,%
        bbllx=30pt,bblly=350pt,bburx=520pt,bbury=650pt,%
        scale=0.90}
\vspace*{1.4cm}
\caption{\it Branching ratios of the charged Higgs boson within the
$M_h^{125}$ scenario as a function of the charged Higgs mass for two
values of $\tgb=10,40$. The results shown here include the results of this
paper. The branching ratios into charginos and neutralinos
($\tilde\chi\tilde\chi$) involve the sum over all gaugino final states,
whereas those into squarks ($\tilde q\tilde q$) incorporate the sum over
all squark final states. This figure has been obtained with {\tt Hdecay}
\cite{hdecay}.}
\label{fg:brc}
\end{figure}

The branching ratios of the charged Higgs boson are shown in
Fig.~\ref{fg:brc} as a function of the charged Higgs mass, for two values
of $\tgb=10,40$. These plots already include the results of our work.
The decay mode into $\tau^+ \nu_\tau$ reaches branching ratios of more
than 90\% below the $t\bar b$ threshold and the muonic one reaches a few
$10^{-3}$. All other leptonic decay channels of the charged Higgs bosons
are unimportant, while decays into a charm plus bottom or strange quark
appear at the percent- or few permille-level. For large charged Higgs
masses beyond the chargino, neutralino and squark thresholds, decays
into these supersymmetric final states are significant, reaching
branching ratios of up to $\sim 80\%$. Below the $t\bar b$ decay threshold,
charged-Higgs decays into off-shell top quarks, $H^+\to t^*\bar b \to
b\bar b W^+$, are relevant \cite{1OFF}. Their branching ratio can reach
the percent level for charged Higgs masses below the top-bottom
threshold.

This article is structured as follows: In Section \ref{sc:yukawa} we
discuss the effective bottom and top Yukawa couplings at NNLO, in
Section \ref{sc:decay} we state our main results for the charged Higgs
decays $H^+\to t\bar b, c\bar b, c\bar s$ before concluding in Section
\ref{sc:conclusions}.

\section{Effective Yukawa couplings} \label{sc:yukawa}
Within the MSSM, radiative corrections to the effective bottom and top
Yukawa couplings are important for moderate to large values of $\tgb$.
The dominant part of these corrections can be coped with by introducing
effective Yukawa factors for the neutral Higgs bosons. For charged Higgs
bosons, the top and bottom Yukawa-coupling factors are identical to the
pseudoscalar couplings at LO. In this work, we have also adopted the
effective pseudoscalar Yukawa-coupling factors for the charged Higgs boson
that include higher-order corrections. The expressions
for the effective pseudoscalar Yukawa-coupling factors are
\begin{eqnarray}
g_b^A \to \tilde g^A_b & = & \frac{g^A_b}{1+\Delta_b}\left[ 1 -
\frac{\Delta_b}{\mbox{tg}^2\beta} \right] \qquad\qquad (g_b^A = \tgb)
\nonumber \\
g_t^A \to \tilde g^A_t & = & \frac{g^A_t}{1+\Delta_t}\Big[ 1 -
\Delta_t \mbox{tg}^2\beta \Big] \qquad\qquad (g_t^A = 1/\tgb) \, .
\label{eq:gtilde}
\end{eqnarray}
The QCD and electroweak corrections to the bottom Yukawa coupling
\cite{deltab,deltab1} take the form ($C_F = 4/3$)
\begin{eqnarray}
\Delta_b & = & \Delta_b^{QCD} + \Delta_b^{elw,t} + \Delta_b^{elw,1} +
\Delta_b^{elw,2} \nonumber \\
\Delta_b^{QCD} & = & \frac{C_F}{2}~\frac{\alpha_s}{\pi}~m_{\tilde
g}~\mu~\mbox{tg}\beta~ I(m^2_{\tilde{b}_1},m^2_{\tilde{b}_2},m^2_{\tilde
g}) \nonumber \\
\Delta_b^{elw,t} & = &
\frac{\lambda_t^2}{(4\pi)^2}~A_t~\mu~\mbox{tg}\beta~
I(m_{\tilde{t}_1}^2,m_{\tilde{t}_2}^2,\mu^2) \nonumber \\
\Delta_b^{elw,1} & = &
-\frac{\alpha_1}{12\pi}~M_1~\mu~\mbox{tg}\beta~\left\{
\frac{1}{3}I(m_{\tilde b_1}^2,m_{\tilde b_2}^2,M_1^2)
+\left( \frac{c_b^2}{2}+s_b^2\right)I(m_{\tilde b_1}^2,M_1^2,\mu^2)
\right. \nonumber \\
& & \left. \hspace*{4cm} +\left( \frac{s_b^2}{2}+c_b^2\right)
I(m_{\tilde b_2}^2,M_1^2,\mu^2)
\right\} \nonumber \\
\Delta_b^{elw,2} & = &
-\frac{\alpha_2}{4\pi}~M_2~\mu~\mbox{tg}\beta~\left\{
c_t^2 I(m_{\tilde t_1}^2,M_2^2,\mu^2)
+ s_t^2 I(m_{\tilde t_2}^2,M_2^2,\mu^2)
\right. \nonumber \\
& & \left. \hspace*{4cm} +\frac{c_b^2}{2} I(m_{\tilde
b_1}^2,M_2^2,\mu^2)
+\frac{s_b^2}{2} I(m_{\tilde b_2}^2,M_2^2,\mu^2)
\right\} \, ,
\label{eq:deltab}
\end{eqnarray}
where $\lambda_t = \sqrt{2} m_t/(v \sin\beta)$ represents the top Yukawa
coupling and $\alpha_1 = {g'}^2/4\pi$, $\alpha_2 = {g}^2/4\pi$
correspond to the electroweak gauge couplings. The masses $m_{{\tilde
b}_{1,2}}$ and $m_{{\tilde t}_{1,2}}$ denote the sbottom and stop
masses. The variables $s/c_{t,b} = \sin/\cos \theta_{t,b}$ are related
to the stop/sbottom mixing angles $\theta_{t,b}$. The function $I$ is
defined as
\begin{equation}
I(a,b,c) = \frac{\displaystyle ab\log\frac{a}{b} + bc\log\frac{b}{c}
+ ca\log\frac{c}{a}}{(a-b)(b-c)(a-c)} = \frac{1}{a-b} \left\{
\frac{\displaystyle a\log\frac{a}{c}}{a-c} - \frac{\displaystyle
b\log\frac{b}{c}}{b-c} \right\} \, .
\label{eq:ii}
\end{equation}
As can be seen from Eq.~(\ref{eq:deltab}),the $\Delta_b$ terms grow with
$\tgb$ in addition to the $\tgb$ behaviour of the bottom Yukawa coupling
at LO. The effective coupling $\tilde g_b^A$ resums the $\Delta_b$
contributions to all orders. The two-loop SUSY--QCD corrections to all
individual terms given in Eq.~(\ref{eq:deltab}) are known
\cite{deltabnnlo, deltabnnlo1, deltabnnlo2}.  They modify the effective
Yukawa couplings by about 10\% for the central renormalization-scale
choice given by the average mass of the contributing SUSY particles at
one-loop order. Potentially large terms growing with $A_b$ can be
incorporated as well by the simple replacement \cite{deltab1}
\begin{eqnarray}
\Delta_b & \to & \frac{\Delta_b}{1+\Delta_{b,1}} \, , \nonumber \\
\mbox{where} \quad \Delta_{b,1} & = &
-\frac{2}{3}~\frac{\alpha_s}{\pi}~m_{\tilde g}~A_b~
I(m^2_{\tilde{b}_1},m^2_{\tilde{b}_2},m^2_{\tilde g}) \, .
\end{eqnarray}
Here the two-loop SUSY--QCD corrections amount to $\sim 10\%$
\cite{deltabnnlo2}. These one- and two-loop calculations can be
translated to the strange Yukawa coupling with the appropriate
replacements of the bottom/top masses and couplings by their
strange/charm counter parts \cite{deltabnnlo2}.

\begin{figure}[htbp]
\begin{center}
\vspace*{-3.0cm}
\epsfig{file=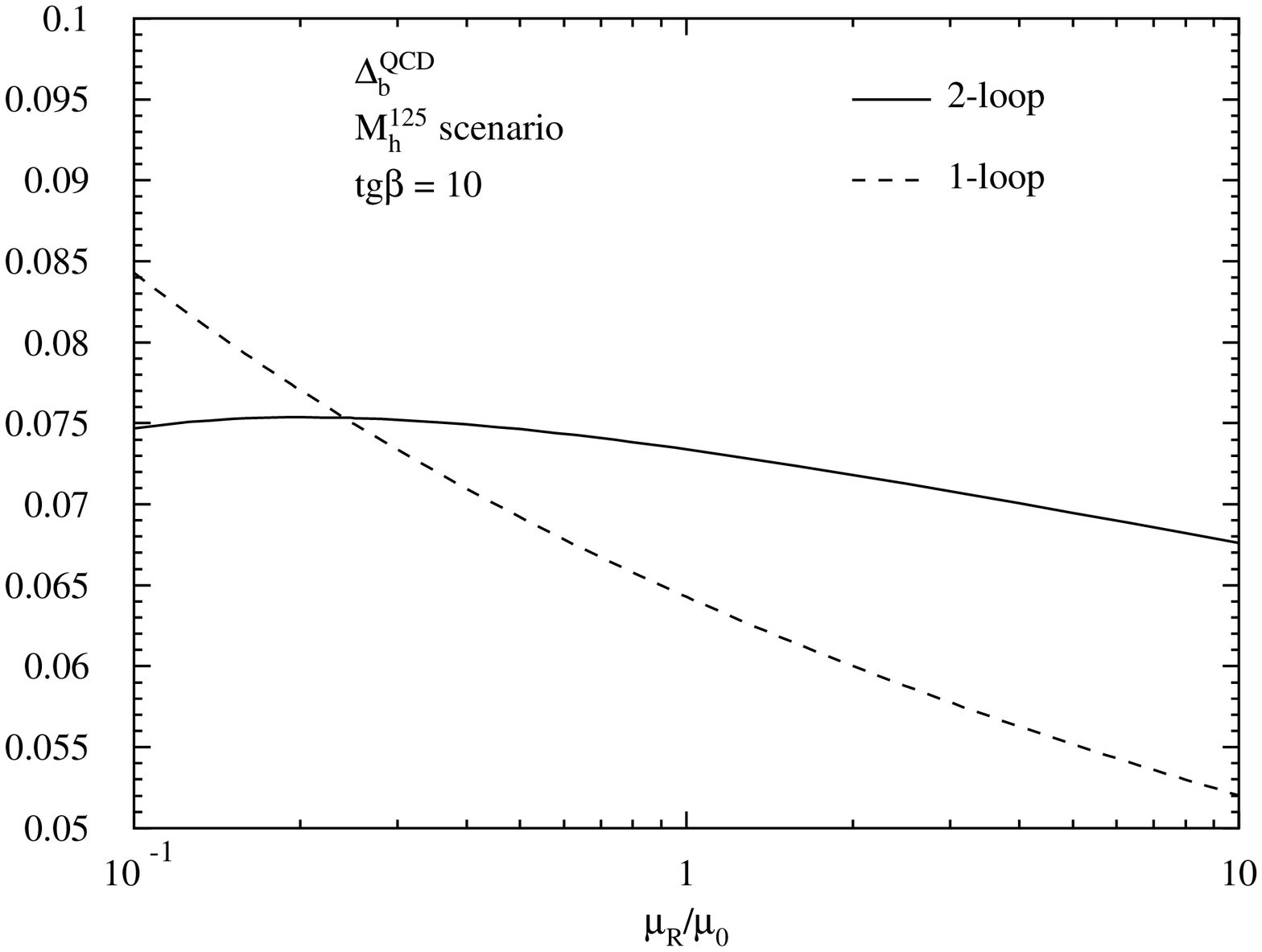,%
        bbllx=30pt,bblly=350pt,bburx=520pt,bbury=650pt,%
        scale=0.6}
\vspace*{2.5cm}

\epsfig{file=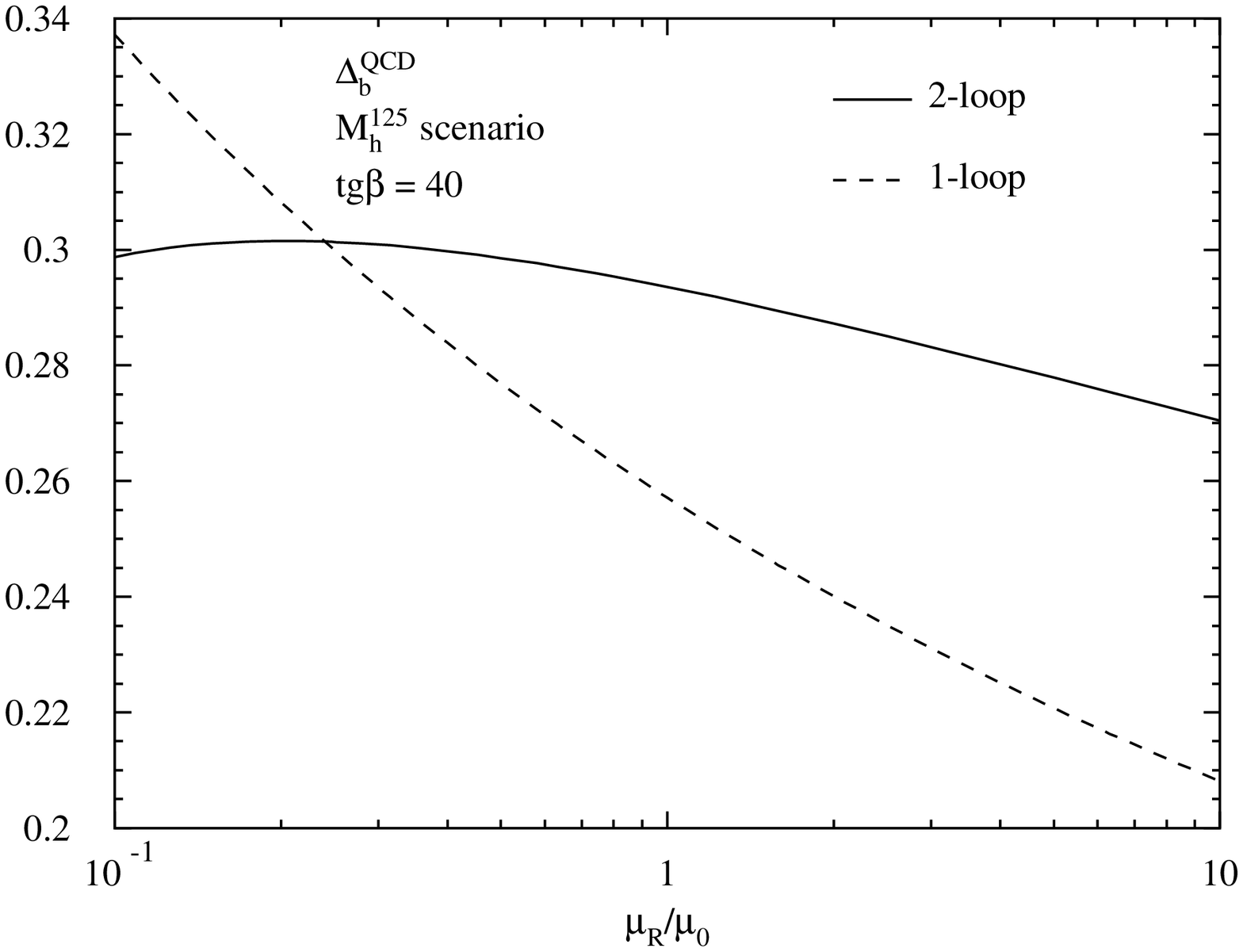,%
        bbllx=30pt,bblly=350pt,bburx=520pt,bbury=650pt,%
        scale=0.6}
\end{center}
\vspace*{2.0cm}
\caption{\it Scale dependence of the SUSY--QCD correction
$\Delta_b^{QCD}$ given in Eq.~(\ref{eq:deltab}) at one-loop and two-loop
order in the $M_h^{125}$ scenario for $\tgb=10,40$ as a function of the
renormalization scale $\mu_R$ and in units of the central scale choice
$\mu_0 = (m_{\tilde{b}_1} + m_{\tilde{b}_2} + m_{\sgl})/3$.}
\label{fg:delta_b}
\end{figure}

For our calculation of the charged Higgs decays into quarks, the
SUSY--QCD correction $\Delta_b^{QCD}$ is the relevant one for analyzing
the different perturbative orders in the strong coupling $\alpha_s$. The
SUSY--QCD part $\Delta_b^{QCD}$ at one- and two-loop order is displayed
in Fig.~\ref{fg:delta_b} as a function of the renormalization scale of
the strong coupling $\alpha_s$.

\begin{figure}[htbp]
\begin{center}
\vspace*{-2.0cm}
\epsfig{file=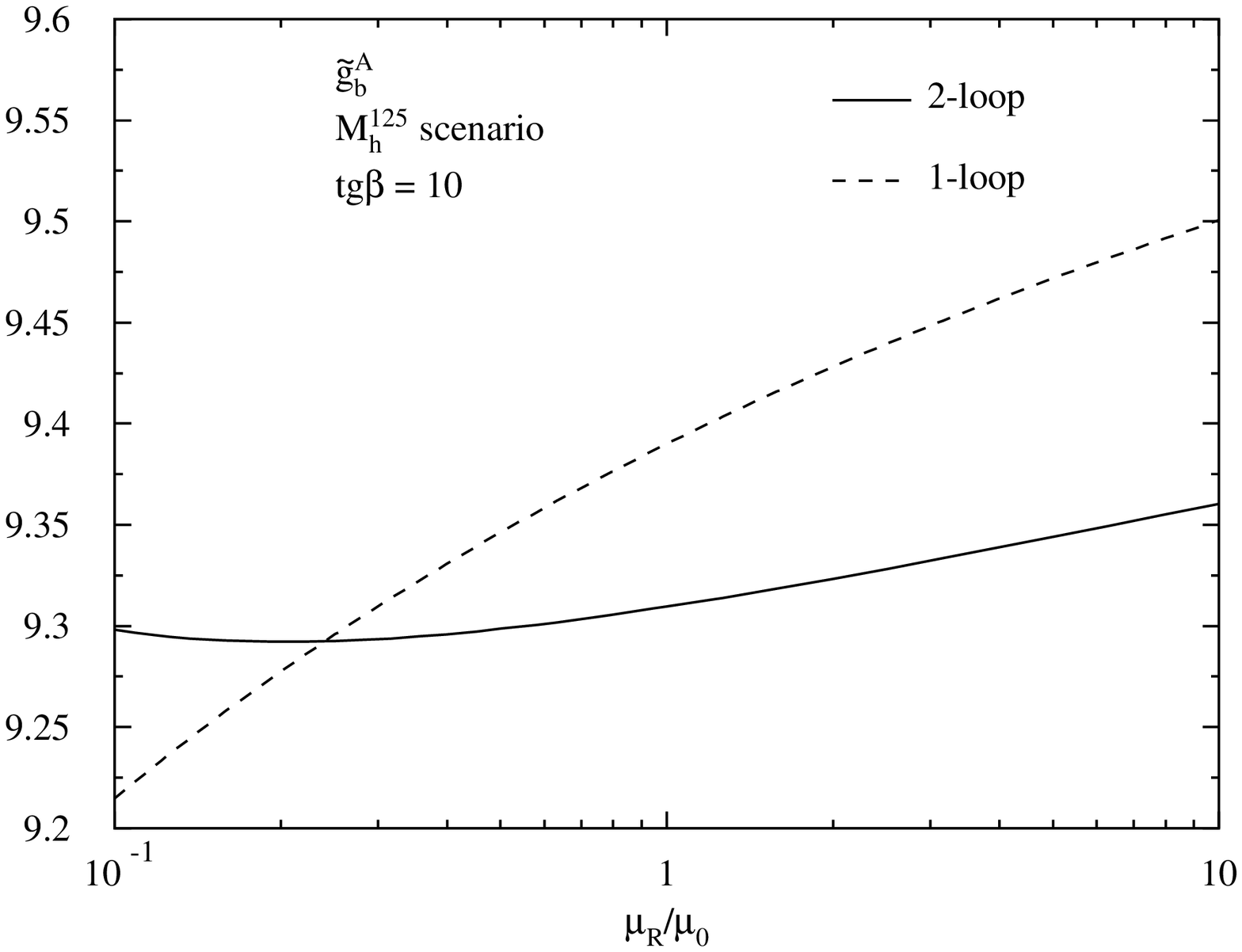,%
        bbllx=30pt,bblly=350pt,bburx=520pt,bbury=650pt,%
        scale=0.6}
\vspace*{2.5cm}

\epsfig{file=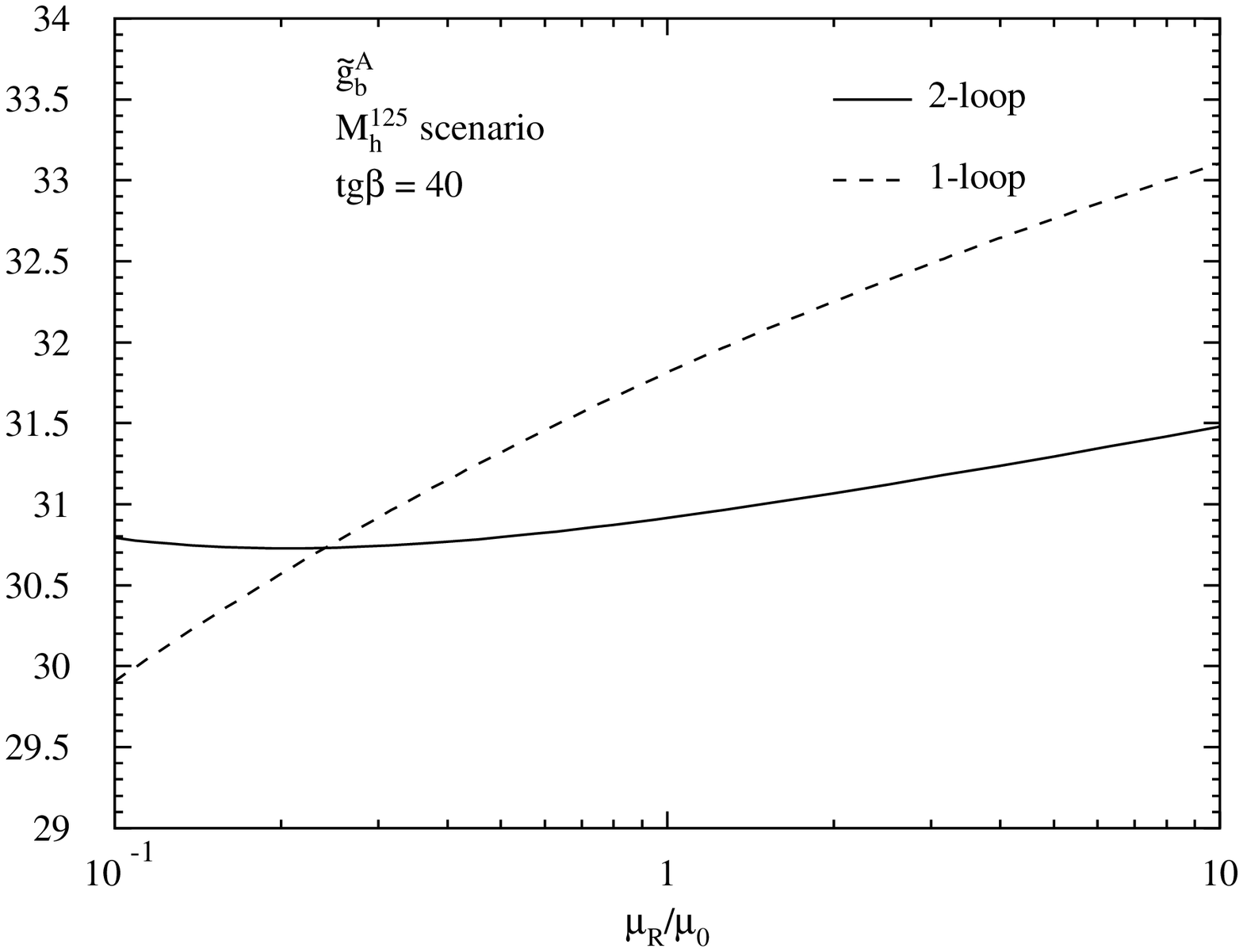,%
        bbllx=30pt,bblly=350pt,bburx=520pt,bbury=650pt,%
        scale=0.6}
\end{center}
\vspace*{2.0cm}
\caption{\it Scale dependence of the SUSY--QCD corrected effective
Yukawa coupling factors $\tilde g_b^A$ of Eq.~(\ref{eq:gtilde}) at
one-loop and two-loop order in the $M_h^{125}$ scenario as a function of
the renormalization scale $\mu_R$ and in units of the central scale
choice $\mu_0 = (m_{\tilde{b}_1} + m_{\tilde{b}_2} + m_{\sgl})/3$. At
leading order, $g_b^A = 10 (40)$ for $\tgb=10 (40)$.}
\label{fg:g_b}
\end{figure}

The $\Delta_b$ corrections to the bottom Yukawa coupling amount to up to
$30\%-40\%$, depending on the value of $\tgb$ in the $M_h^{125}$
benchmark scenario, and are thus sizable. The scale dependence is
significantly reduced from one- to two-loop order so that the residual
theoretical uncertainties remain at the few-percent level. For the
central prediction of the $\Delta_b^{QCD}$ corrections to the bottom
Yukawa coupling, we have chosen the average SUSY mass $\mu_R =
(m_{\tilde{b}_1} + m_{\tilde{b}_2} + m_{\sgl})/3$ as the renormalization
scale. The corresponding effective Yukawa-coupling factor $\tilde g_b^A$
of Eq.~(\ref{eq:gtilde}) is displayed in Fig.~\ref{fg:g_b} as a function
of the renormalization scale $\mu_R$.
The radiative corrections included in $\Delta_b$ reduce the LO coupling
factor $g_b^A$ by about 10\% (20\%) for $\tgb=10 (40)$, as can be
inferred from the comparison of the effective $\tilde g_b^A$ factor with
the LO coupling $g_b^A$ for the central scale choice $\mu_R=\mu_0$.

\subsection{Effective top Yukawa couplings at NNLO} \label{sc:t-yukawa}
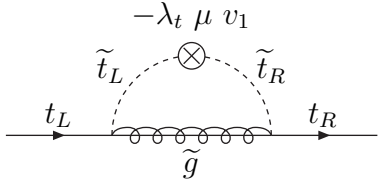
\begin{figure}[htbp]
\begin{center}
\begin{picture}(300,60)(-80,0)
\Quark{0}{0}{40}{0}
\Gluino{40}{0}{100}{0}{6}
\Quark{100}{0}{140}{0}
\DashCArc(70,0)(30,0,180){2}
\CrossedCircle{70}{30}{67}{27}{73}{33} 
\Text(20,3)[cb]{$\tL$}
\Text(120,3)[cb]{$\tR$}
\Text(45,20)[rb]{$\tsL$}
\Text(95,20)[lb]{$\tsR$} 
\Text(70,-5)[ct]{$\gluino$}
\Text(70,40)[cb]{$-\lt\ \mu\ v_1$}
\end{picture}
\end{center}
\caption{\it One-loop diagrams of the SUSY--QCD contributions to the top
self-energy with the off-diagonal mass insertions corresponding to the
corrections $\Delta_t$ of the top Yukawa couplings. The contributing
particles involve top quarks $t$, top squarks $\st$ and gluinos
$\gluino$.}
\label{fg:delta_t1}
\end{figure}

The top Yukawa coupling is affected by analogous radiative corrections.
The SUSY--QCD part is discussed at the NNLO level in this
section\footnote{The two-loop corrections to the effective top Yukawa
coupling of the light scalar Higgs boson $h$ have first been calculated
in Ref.~\cite{mihailazerf}.}. At the one-loop level, it is given by
\begin{eqnarray}
\Delta_t & = & \frac{C_F}{2}~\frac{\alpha_s}{\pi}~\frac{m_{\tilde
g}~\mu}{\mbox{tg}\beta}~
I(m^2_{\tilde{t}_1},m^2_{\tilde{t}_2},m^2_{\tilde g}) \, .
\label{eq:deltat}
\end{eqnarray}
This term is the leading SUSY-QCD correction to the top Yukawa couplings
of the {\it neutral} MSSM Higgs in the limit of heavy SUSY particles,
analogously to the bottom case \cite{deltab1}. For the incorporation of
the NNLO corrections, we follow the same line as in
Refs.~\cite{deltab1,deltabnnlo}. The one-loop corrections can be
obtained by off-diagonal mass insertions as shown in
Fig.~\ref{fg:delta_t1} and by replacing the vacuum expectation value
$v_1$ of the first Higgs doublet by the full Higgs field, $v_1
\to \sqrt{2}\phi_1^{0*}$. This method is based on the low-energy
theorems for soft external Higgs fields \cite{let}
and results in the effective Lagrangian for the neutral Higgs fields
\begin{eqnarray}
{\cal L}_{eff} & = & -\lambda_t \overline{t_R} \left[ \phi_2^0
+ \Delta_t \tgb \phi_1^{0*} \right] t_L + h.c. \nonumber \\
& = & -m_t \bar t \left[1-i\gamma_5 \frac{G^0}{v}\right] t
-\frac{m_t/v}{1+\Delta_t} \bar t \left[ g_t^h \left(
1-\Delta_t~\tga~\tgb\right) h \right. \nonumber \\
& & \hspace*{2cm} \left. + g_t^H \left( 1+\Delta_t
\frac{\tgb}{\tga}\right) H
- g_t^A \left(1-\Delta_t \stgb \right) i \gamma_5 A \right] t \, ,
\label{eq:leff}
\end{eqnarray}
where $g_{t,b}^A$ are as defined in Eq.~(\ref{eq:gtilde}) and we
used the explicit representation of the neutral components of the Higgs
iso-doublets,
\begin{eqnarray}
\phi_1^0 & = & \frac{1}{\sqrt{2}}\left[ v_1 + H\coa - h\sia + iA\sib -
iG^0\cob
\right] \nonumber \\
\phi_2^0 & = & \frac{1}{\sqrt{2}}\left[ v_2 + H\sia + h\coa + iA\cob +
iG^0\sib
\right] \, .
\end{eqnarray}
In these expressions, $\alpha$ denotes the mixing angle of the
neutral CP-even Higgs sector. The relation between the top Yukawa coupling
$\lambda_t$ and the top mass $m_t$ is modified as
\begin{equation}
\lambda_t = \sqrt{2} \frac{m_t}{v \sin \beta (1+\Delta_t)} \, .
\end{equation}
In this work, we assume the pseudoscalar coupling $g_t^A$ to top quarks
to be the relevant effective top-Yukawa coupling for the charged Higgs
boson, also at higher perturbative order. Power counting allows
us to show that the effective Lagrangian on the first line of
Eq.~(\ref{eq:leff}), which is given in the current-eigenstate basis, is
exact for the leading terms scaling with $1/\tgb$.  Any additional
insertion of the vertex Feynman rule of Fig.~\ref{fg:delta_t1} yields
another power of $m_t/M_{SUSY}$.  The leading term of the full vertex in
the expansion in $m_t^2/M_{SUSY}^2$ and $M_\Phi^2/M_{SUSY}^2$
$(\Phi=h,H,A)$ is given by the contribution $\Delta_t$ in the limit of
vanishing top and Higgs masses.

\begin{figure}[htbp]
\SetScale{0.8}
\begin{picture}(200,40)(-20,0)
\ArrowLine(0,0)(50,0)
\Line(50,0)(150,0)
\Gluon(50,0)(100,0){5}{3}
\Gluon(100,0)(150,0){5}{3}
\Gluon(100,50)(100,0){5}{3}
\ArrowLine(150,0)(200,0)
\DashCArc(100,0)(50,0,180){5}
\put(10,6){$t$}
\put(148,6){$t$}
\put(88,16){$g$}
\put(56,-15){$\sgl$}
\put(100,-15){$\sgl$}
\put(38,25){$\st$}
\put(116,25){$\st$}
\end{picture}
\begin{picture}(200,40)(-10,0)
\ArrowLine(0,0)(50,0)
\Line(50,0)(100,0) 
\Gluon(50,0)(100,0){5}{3}
\ArrowLine(100,0)(150,0)
\DashLine(100,50)(100,0){5}
\ArrowLine(150,0)(200,0) 
\GlueArc(100,0)(50,0,90){5}{5}
\DashCArc(100,0)(50,90,180){5}
\put(10,6){$t$}
\put(148,6){$t$}
\put(86,16){$\st$}
\put(56,-15){$\sgl$}
\put(100,-12){$t$}
\put(44,32){$\st$}
\put(116,32){$g$}
\end{picture} \\
\begin{picture}(200,90)(-20,0)
\ArrowLine(0,0)(50,0) 
\ArrowLine(50,0)(100,0)
\Line(100,0)(150,0)
\Gluon(100,0)(150,0){5}{3}
\DashLine(100,50)(100,0){5}
\ArrowLine(150,0)(200,0) 
\GlueArc(100,0)(50,90,180){5}{5}
\DashCArc(100,0)(50,0,90){5}
\put(10,6){$t$}
\put(148,6){$t$}
\put(86,16){$\st$}
\put(56,-12){$t$}
\put(100,-15){$\sgl$}
\put(40,32){$g$}
\put(116,32){$\st$}
\end{picture}
\begin{picture}(200,90)(-10,0)
\ArrowLine(0,0)(50,0)
\Line(50,0)(100,0) 
\Line(100,50)(100,0) 
\Gluon(50,0)(100,0){5}{3}
\Gluon(100,0)(150,0){5}{3}
\Gluon(100,50)(100,0){5}{3}
\ArrowLine(150,0)(200,0)
\DashCArc(100,0)(50,90,180){5}
\ArrowArcn(100,0)(50,90,0)
\put(10,6){$t$}
\put(148,6){$t$}
\put(88,16){$\sgl$}
\put(56,-15){$\sgl$}
\put(100,-15){$g$}
\put(38,25){$\st$}
\put(116,25){$t$}
\end{picture} \\
\begin{picture}(200,90)(-10,0)
\ArrowLine(0,0)(50,0)
\Line(100,0)(150,0)
\Line(100,50)(100,0)
\Gluon(50,0)(100,0){5}{3}
\Gluon(100,0)(150,0){5}{3}
\Gluon(100,50)(100,0){5}{3}
\ArrowLine(150,0)(200,0)
\DashCArc(100,0)(50,0,90){5}
\ArrowArcn(100,0)(50,180,90)
\put(10,6){$t$}
\put(148,6){$t$}
\put(88,16){$\sgl$}
\put(56,-15){$g$}
\put(100,-15){$\sgl$}
\put(38,25){$t$}
\put(116,25){$\st$}
\end{picture}
\begin{picture}(200,90)(-20,0)
\ArrowLine(0,0)(50,0)
\Line(50,0)(100,0)
\Gluon(50,0)(100,0){5}{3}
\DashLine(100,0)(150,0){5}
\ArrowLine(100,50)(100,0)
\ArrowLine(150,0)(200,0)
\CArc(100,0)(50,0,90)
\GlueArc(100,0)(50,0,90){5}{5}
\DashCArc(100,0)(50,90,180){5}
\put(10,6){$t$}
\put(148,6){$t$}
\put(86,16){$t$}
\put(56,-15){$\sgl$}
\put(100,-15){$\st$}
\put(44,32){$\st$}
\put(116,32){$\sgl$}
\end{picture} \\
\begin{picture}(200,90)(-10,0)
\ArrowLine(0,0)(50,0)
\ArrowLine(150,0)(200,0)
\ArrowLine(50,0)(70,0)
\ArrowLine(130,0)(150,0)
\Line(70,0)(130,0)
\Gluon(70,0)(130,0){5}{4}
\GlueArc(100,0)(50,0,180){5}{10}
\DashCArc(100,0)(30,0,180){5}
\put(10,6){$t$}
\put(148,6){$t$}
\put(48,-12){$t$}
\put(76,-16){$\sgl$}
\put(112,-12){$t$}
\put(76,12){$\st$}
\put(116,34){$g$}
\end{picture}
\begin{picture}(200,90)(-10,0)
\ArrowLine(0,0)(50,0)
\Line(50,0)(150,0)
\Gluon(50,0)(150,0){5}{7}
\ArrowLine(150,0)(200,0)
\DashCArc(100,0)(50,0,180){5}
\GOval(100,46)(10,20)(0){0.5}
\put(10,6){$t$}
\put(148,6){$t$}
\put(72,-15){$\sgl$}
\put(36,25){$\st$}
\put(116,25){$\st$}
\end{picture} \\
\begin{picture}(200,90)(-10,0)
\ArrowLine(0,0)(50,0)
\Line(50,0)(150,0)
\Gluon(50,0)(150,0){5}{7}
\ArrowLine(150,0)(200,0)
\DashCArc(100,0)(50,0,180){5}
\GlueArc(100,3.5)(25,0,180){5}{5}
\put(10,6){$t$}
\put(148,6){$t$}
\put(47,-14){$\sgl$}
\put(75,-14){$\sgl$}
\put(107,-14){$\sgl$}
\put(76,30){$g$}
\put(44,32){$\st$}
\end{picture}
\begin{picture}(200,90)(-10,0)
\ArrowLine(0,0)(50,0)
\Line(50,0)(150,0)
\Gluon(50,0)(80,0){5}{2}
\ArrowLine(80,0)(120,0)
\DashCArc(100,0)(20,0,180){5}
\Gluon(120,0)(150,0){5}{2}
\ArrowLine(150,0)(200,0)
\DashCArc(100,0)(50,0,180){5}
\put(10,6){$t$}
\put(148,6){$t$}
\put(50,-14){$\sgl$}
\put(105,-14){$\sgl$}
\put(75,-12){$q$}
\put(75,20){$\tilde q$}
\put(44,32){$\st$}
\end{picture} \\
\begin{picture}(200,70)(-10,0)
\DashLine(0,0)(100,0){5}
\GOval(50,0)(10,20)(0){0.5}
\DashLine(130,0)(230,0){5}
\GlueArc(180,0)(20,0,180){4}{5}
\DashLine(260,0)(360,0){5}
\ArrowLine(290,0)(330,0)
\CArc(310,0)(20,0,180)
\GlueArc(310,0)(20,0,180){4}{5}
\DashLine(390,0)(490,0){5}
\DashCArc(440,20)(20,0,360){5}
\Vertex(440,0){2}
\put(88,-3){$=$}
\put(193,-3){$+$}
\put(297,-3){$+$}
\put(10,6){$\st$}
\put(70,6){$\st$}
\put(113,6){$\st$}
\put(143,-13){$\st$}
\put(143,25){$g$}
\put(175,6){$\st$}
\put(218,6){$\st$}
\put(248,-12){$t$}
\put(248,25){$\sgl$}
\put(280,6){$\st$}
\put(323,6){$\st$}
\put(351,35){$\st$}
\put(383,6){$\st$}
\end{picture} \\
\caption{\it Two-loop diagrams of the SUSY--QCD contributions to the top
self-energy involving top quarks $t$, top squarks $\st$, gluons $g$ and
gluinos $\sgl$. The squark-quark contributions to the gluino propagator
have to be summed over all quark/squark flavors $q/\tilde q$, including
both directions of the flavor flow due to the Majorana nature of the
gluino.}
\label{fg:2-loop}
\end{figure}
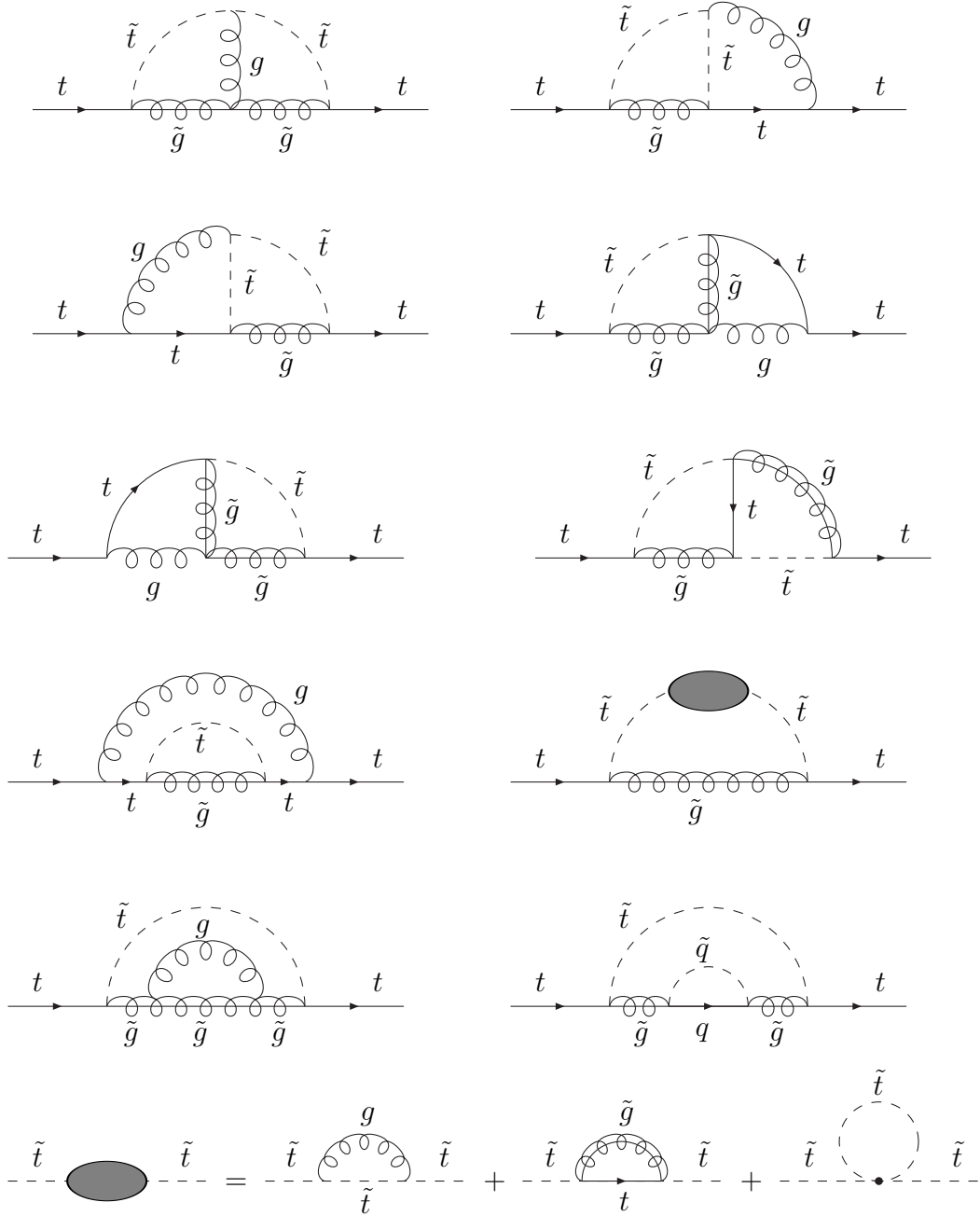

The two-loop corrections to $\Delta_t$ can be derived from the related
two-loop corrections to the top-quark self-energy (see
Fig.~\ref{fg:2-loop}) in the limit of vanishing quark masses (including
the top mass) for the first term of a large SUSY-mass expansion.
Applying all possible off-diagonal mass insertions in the stop
propagators of Fig.~\ref{fg:2-loop}, we have obtained the two-loop
corrections to the $\Delta_t$ term. The stop and gluino masses have been
renormalized on-shell. For the strong coupling $\alpha_s$ we have
applied the $\overline{\rm MS}$ scheme with five active flavours, where
the top quark and SUSY particles have been decoupled. The explicit
expressions of the renormalization constants have been obtained by
replacing the sbottom masses by the stop masses in the results given in
Ref.~\cite{deltabnnlo}. In order to restore the supersymmetric
relations between the SM couplings and their supersymmetric
counterparts, we have included the anomalous counterterms within
dimensional regularization which we used in the calculation of the NNLO
corrections to the top Yukawa coupling. In this way, we have translated
the calculation of the NNLO corrections presented for the bottom-quark
case in Ref.~\cite{deltabnnlo} to the top-quark case.
\begin{figure}[htbp]
\begin{center}
\vspace*{-1.0cm}
\epsfig{file=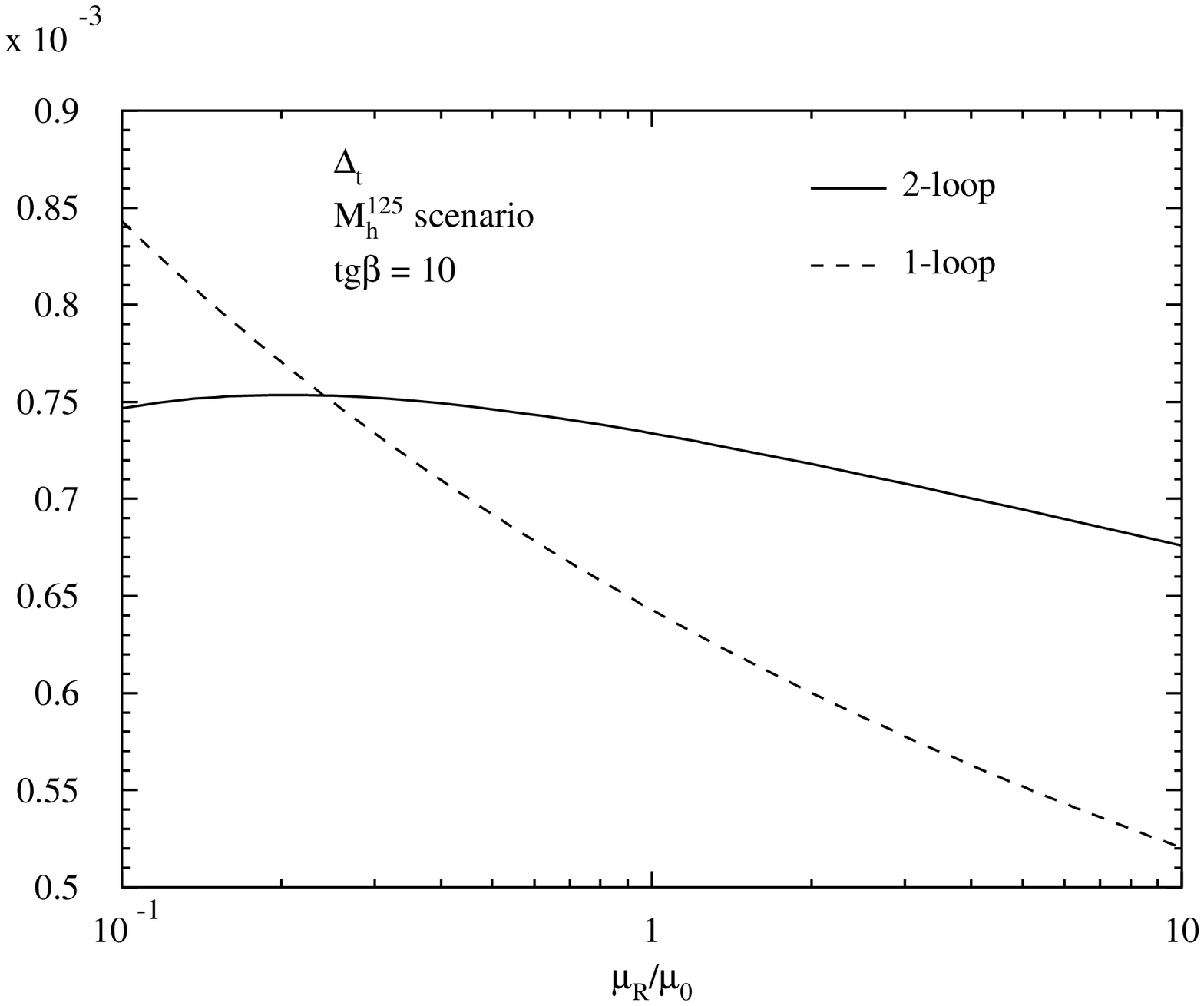,%
        bbllx=30pt,bblly=350pt,bburx=520pt,bbury=650pt,%
        scale=0.6}
\vspace*{2.5cm}

\epsfig{file=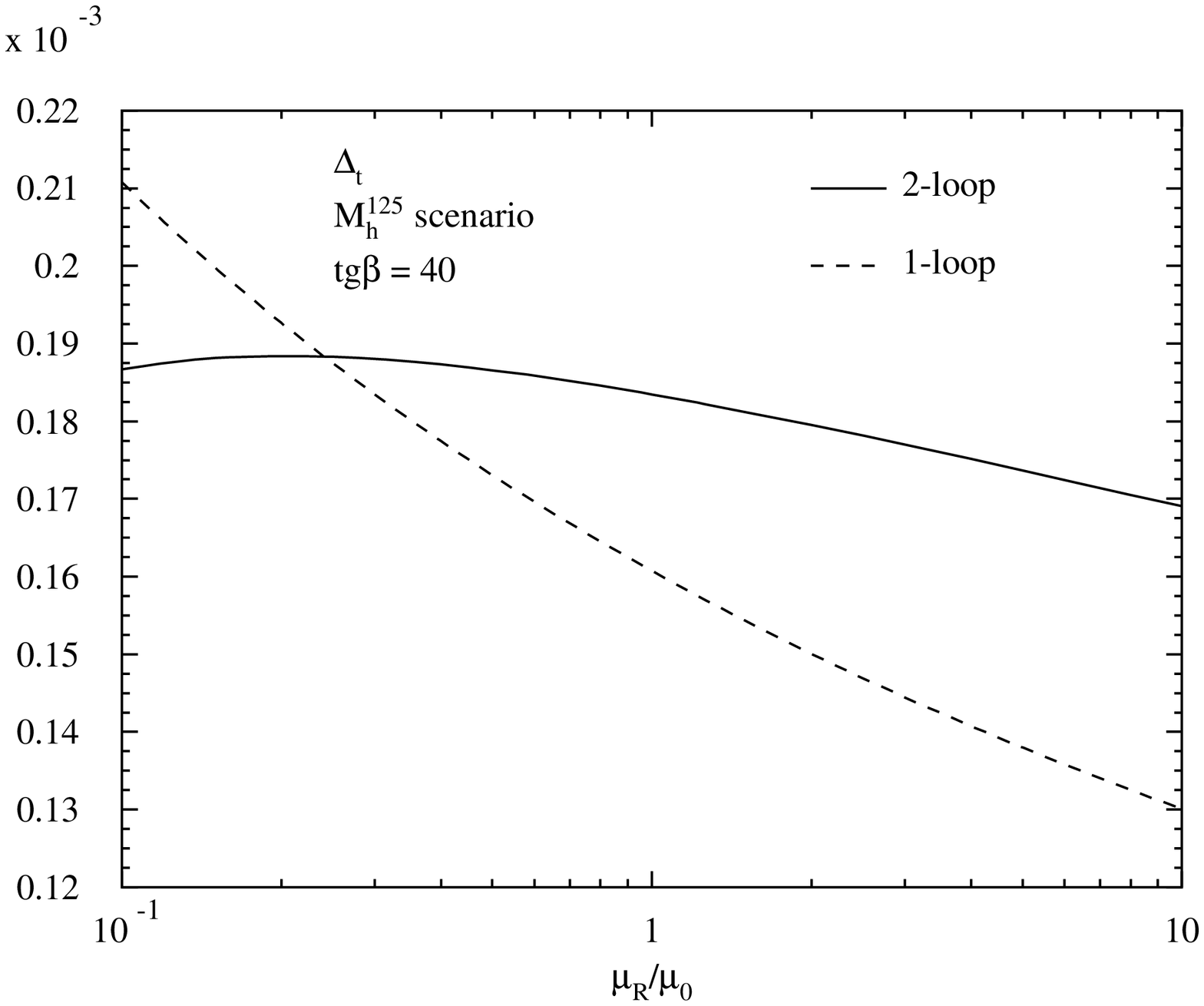,%
        bbllx=30pt,bblly=350pt,bburx=520pt,bbury=650pt,%
        scale=0.6}
\end{center}
\vspace*{2.0cm}
\caption{\it Scale dependence of the SUSY--QCD correction $\Delta_t$ at
one-loop and two-loop order in the $M_h^{125}$ scenario for $\tgb=10,40$
as a function of the renormalization scale $\mu_R$ and in units of the
central scale choice $\mu_0 = (m_{\tilde{t}_1} + m_{\tilde{t}_2} +
m_{\sgl})/3$.}
\label{fg:delta_t}
\end{figure}
\begin{figure}[htbp]
\begin{center}
\vspace*{-3.0cm}
\epsfig{file=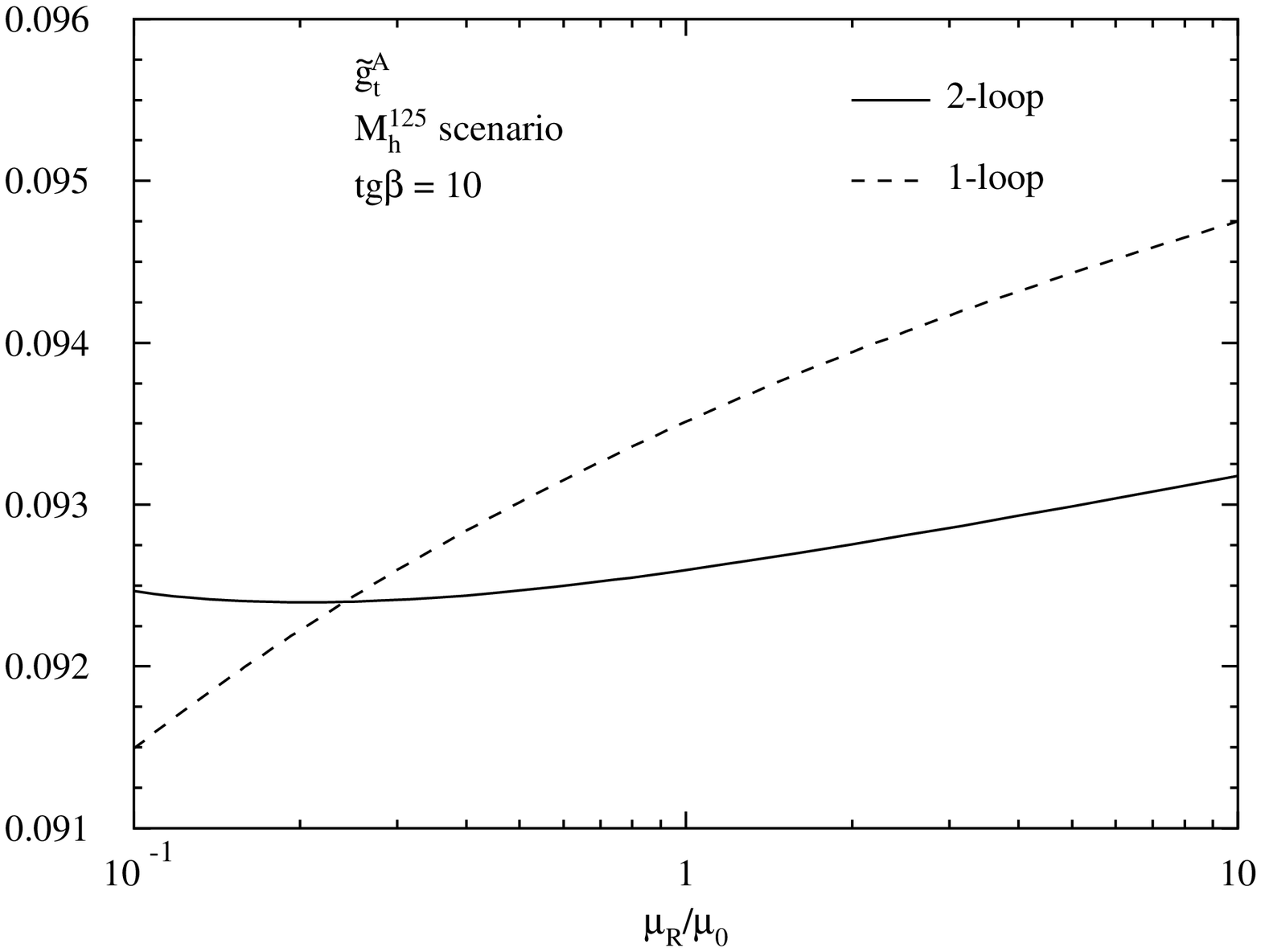,%
        bbllx=30pt,bblly=350pt,bburx=520pt,bbury=650pt,%
        scale=0.6}
\vspace*{2.5cm}

\epsfig{file=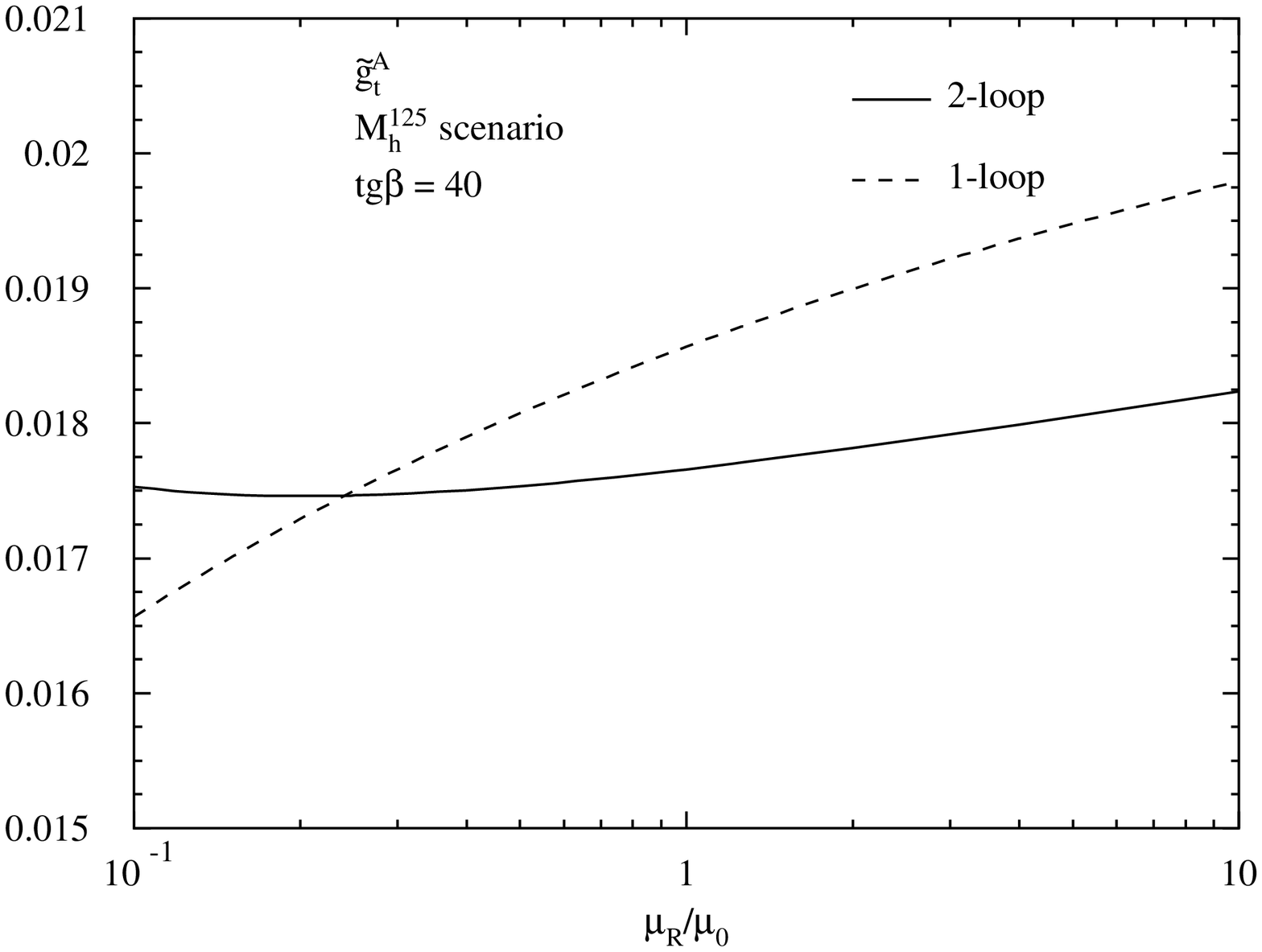,%
        bbllx=30pt,bblly=350pt,bburx=520pt,bbury=650pt,%
        scale=0.6}
\end{center}
\vspace*{2.0cm}
\caption{\it Scale dependence of the SUSY--QCD corrected effective
Yukawa-coupling factors $\tilde g_t^A$ at one-loop and two-loop order in the
$M_h^{125}$ scenario as a function of the renormalization scale $\mu_R$
and in units of the central scale choice $\mu_0 = (m_{\tilde{t}_1} +
m_{\tilde{t}_2} + m_{\sgl})/3$. At leading
order, $g_b^A = 0.1 (0.025)$ for $\tgb=10 (40)$.}
\label{fg:g_t}
\end{figure}
The results for $\Delta_t$ at one- and two-loop order are shown in
Fig.~\ref{fg:delta_t} as a function of the renormalization scale of the
strong coupling $\alpha_s$. The $\Delta_t$ terms turn out to be at the
permille level in the $M_h^{125}$ benchmark scenario. The scale
dependence is significantly reduced from the one- to two-loop level so
that the residual relative theoretical uncertainties are at the
few-percent level. For the central prediction of the $\Delta_t$
contributions to the effective top Yukawa coupling we have chosen the
average SUSY mass $\mu_R = (m_{\tilde{t}_1} + m_{\tilde{t}_2} +
m_{\sgl})/3$ as the renormalization scale. The corresponding effective
top Yukawa-coupling factor $\tilde g_t^A$ of Eq.~(\ref{eq:gtilde}) is
presented in Fig.~\ref{fg:g_t} at one- and two-loop order as a function
of the renormalization scale $\mu_R$. Due to the additional
$\mbox{tg}^2\beta$ enhancement of $\tilde g_t^A$, the SUSY--QCD
corrections to the effective Yukawa-coupling factor turn out to be
sizable, i.e.~about 5\%-10\% (30\%) for $\tgb = 10 (40)$, and thus of similar
size as the bottom Yukawa-coupling factor $\tilde g_b^A$ discussed in
the previous subsection.

\section{Charged Higgs decays into quarks} \label{sc:decay}
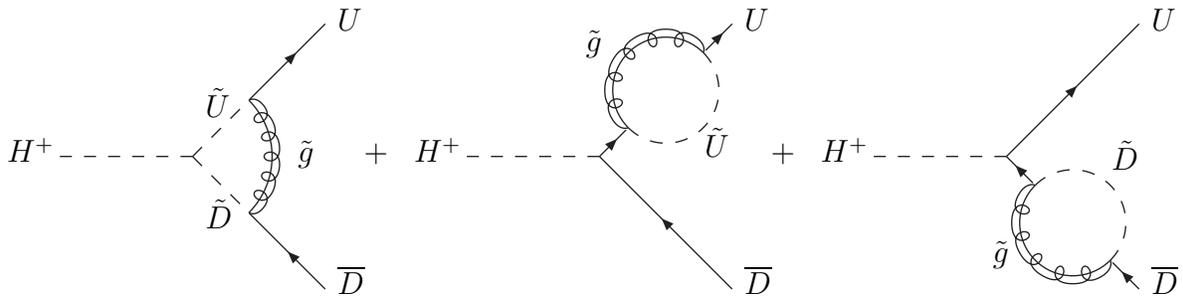
\begin{figure}[htb]
\begin{center}
\begin{picture}(130,95)(0,0)
\DashLine(0,50)(50,50){5}
\ArrowLine(75,75)(100,100)
\ArrowLine(100,0)(75,25)
\DashLine(50,50)(75,75){5}
\CArc(50,50)(30,-45,45)
\GlueArc(50,50)(30,-45,45){3}{5}
\DashLine(75,25)(50,50){5}
\put(-20,48){$H^+$}
\put(105,98){$U$}
\put(105,-2){$\overline{D}$}
\put(55,65){$\tilde U$}
\put(55,23){$\tilde D$}
\put(90,48){$\tilde g$}
\put(115,48){$+$}
\end{picture}
\begin{picture}(130,95)(-20,0)
\DashLine(0,50)(50,50){5}
\ArrowLine(50,50)(60,60)
\ArrowLine(90,90)(100,100)
\ArrowLine(100,0)(50,50)
\CArc(75,75)(20,45,225)
\GlueArc(75,75)(20,45,225){3}{5}
\DashCArc(75,75)(20,-135,45){5}
\put(-20,48){$H^+$}
\put(105,98){$U$}
\put(105,-2){$\overline{D}$}
\put(90,50){$\tilde U$}
\put(45,90){$\tilde g$}
\put(115,48){$+$}
\end{picture}
\begin{picture}(130,95)(-40,0)
\DashLine(0,50)(50,50){5}
\ArrowLine(50,50)(100,100)
\ArrowLine(100,0)(90,10)
\ArrowLine(60,40)(50,50)
\CArc(75,25)(20,135,315)
\GlueArc(75,25)(20,135,315){3}{5}
\DashCArc(75,25)(20,-45,135){5}
\put(-20,48){$H^+$}
\put(105,98){$U$}
\put(105,-2){$\overline{D}$}
\put(90,45){$\tilde D$}
\put(45,10){$\tilde g$}
\end{picture} \\[-0.0cm]
\caption{\label{fg:dianlo} \it Virtual SUSY--QCD corrections to charged
Higgs boson decays into $U\overline{D}$ quark pairs at next-to-leading
order. $U$ denotes up-type and $D$ down-type quarks.}
\end{center}
\end{figure}
\noindent
In this section, we consider the genuine NLO SUSY--QCD corrections
to the charged Higgs decays into quarks. These consist of
vertex-correction terms and self-energy contributions to the external
legs, see Fig.~\ref{fg:dianlo}, which have been combined with their respective
counterterms.
\begin{figure}[htb]
\begin{center}
\begin{picture}(130,105)(0,0)
\DashLine(0,50)(50,50){5}
\DashLine(50,50)(100,100){5}
\DashLine(100,0)(50,50){5}
\ArrowLine(74,74)(76,76)
\ArrowLine(76,24)(74,26)
\put(120,48){$\displaystyle i\sqrt{2}~\frac{G_{ij}}{v}$}
\put(-20,48){$H^+$}
\put(105,98){$\tilde U_i$}
\put(105,-2){$\tilde D_j$}
\end{picture} \\[-0.0cm]
\caption{\label{fg:htbfeyn} \it Feynman rule for the charged Higgs
coupling to up- and down-type squarks, $\tilde U_i$ and $\tilde D_j$
$(ij = L,R~\mbox{or}~1,2)$, respectively.}
\end{center}
\end{figure}
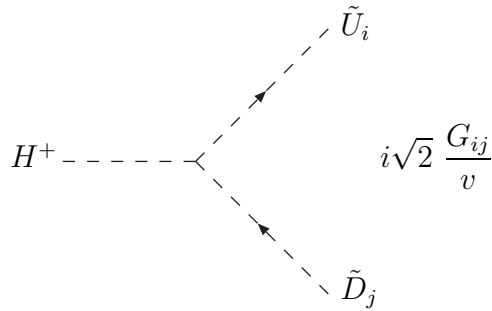
For the calculation of the vertex corrections, the charged Higgs
couplings to up- and down-type squarks are required. The generic Feynman
rule is depicted in Fig.~\ref{fg:htbfeyn}. In the chiral basis, the
corresponding couplings read as ($g_D^A = 1/g_U^A = \tgb$)
\begin{eqnarray}
G_{LL} & = & m_D^2 g_D^A + m_U^2 g_U^A - M_W^2 s_{2\beta} \nonumber \\
G_{LR} & = & m_D (A_D g_D^A + \mu) \nonumber \\
G_{RL} & = & m_U (A_U g_U^A + \mu) \nonumber \\
G_{RR} & = & m_D m_U (g_D^A + g_U^A) \, .
\end{eqnarray}
Rotating these couplings to the sfermion mass eigenstate basis results
in
\begin{eqnarray}
G_{11} & = & G_{LL} c_U c_D + G_{LR} c_U s_D + G_{RL} s_U c_D + G_{RR} s_U s_D
\nonumber \\
G_{12} & = &-G_{LL} c_U s_D + G_{LR} c_U c_D - G_{RL} s_U s_D + G_{RR} s_U c_D
\nonumber \\
G_{21} & = &-G_{LL} s_U c_D - G_{LR} s_U s_D + G_{RL} c_U c_D + G_{RR} c_U s_D
\nonumber \\
G_{22} & = & G_{LL} s_U s_D - G_{LR} s_U c_D - G_{RL} c_U s_D + G_{RR}
c_U c_D \, ,
\end{eqnarray}
where $s_i = \sin \theta_i, c_i = \cos \theta_i$ denote the mixing-angle
contributions.
Since the sfermion mixing angles are proportional to the SM fermion
masses, we have kept the mixing angles for the stop and sbottom states, but
have neglected them for the first two generations.

For the SUSY--QCD part of the NLO calculation, we have renormalized the
quark masses on-shell. The contribution of the SUSY masses to the
QCD-running of the quark masses is decoupled in this way. We have
assumed that the pure QCD corrections factorize from the genuine
SUSY--QCD corrections, since the QCD corrections are dominated by
light-particle contributions. In the QCD part, the up-type (down-type)
quark masses have been implemented as the $\overline{\rm MS}$ masses
$\overline{m}_{U,D} = \overline{m}_{U,D}(M_{H^\pm})$ at the scale of the
charged Higgs mass in the limit of large charged Higgs masses, while
closer to the decay threshold we adopted the pole masses\footnote{Both
regions have been combined by a quartic interpolation w.r.t.~the
masses.} $M_{U,D}$. Close to threshold, the partial decay width at NLO
reads
\begin{eqnarray}
\Gamma [H^+\rightarrow U\bar{D}\,] &=& \frac{3 G_F
M_{H^\pm}}{4\sqrt{2}\pi}
|V_{UD}|^2 \, \lambda^{1/2} \, \Big\{ \nonumber \\
& & \left. (1-\mu_U -\mu_D) \left[
M_U^2 (g_U^A)^2
\left( 1+ \frac{4}{3} \frac{\alpha_s}{\pi} \delta_{UD}^+ \right)
\Big( 1 + \tilde\delta_{UD}^+ \Big) \right. \right.
\nonumber \\
& & \left. \left. \qquad\qquad\qquad\quad
+ M_D^2 (g_D^A)^2 \left( 1+ \frac{4}{3}
\frac{\alpha_s} {\pi} \delta_{DU}^+ \right)
\Big( 1 + \tilde\delta_{DU}^+ \Big)
\right]
\right. \nonumber \\
& & \left. -4M_UM_D \sqrt{\mu_U \mu_D} g_U^A g_D^A
\left( 1+ \frac{4}{3}
\frac{\alpha_s}{\pi} \delta_{UD}^- \right)
\left[ 1 + \tilde\delta_{UD}^- \right] \right\} \, ,
\label{eq:h+udsqcd0}
\end{eqnarray}
with $\mu_i=M_i^2/M_{H^\pm}^2$ and
$\lambda=(1-\mu_U-\mu_D)^2-4\mu_U\mu_D$. The $\tilde \delta_{ij}^\pm$
denote the individual parts of the SUSY--QCD corrections. We have
assumed the CKM matrix element $V_{UD}$ to be the same in the quark and
squark sector so that it factorizes globally. The coupling factors
$g_{Q}^A$ are given in Eq.~(\ref{eq:gtilde}) at LO. The QCD-correction
factors $\delta_{ij}^\pm~~(i,j=U,D)$ read as \cite{hud1,hud}
\begin{eqnarray}
\delta_{ij}^{+} &=&  \frac{9}{4} + \frac{ 3-2\mu_i+2\mu_j}{4} \log
\frac{\mu_i}{\mu_j} + \frac{ (\frac{3}{2}-\mu_i-\mu_j) \lambda+5 \mu_i
\mu_j}{2 \lambda^{1/2} (1-\mu_i -\mu_j)} \log x_i x_j  +B_{ij} \nonumber \\
\delta_{ij}^{-} &=&  3 + \frac{ \mu_j-\mu_i}{2} \log \frac{\mu_i}{\mu_j}
+ \frac{ \lambda +2(1-\mu_i-\mu_j)} { 2 \lambda^{1/2} } \log x_i x_j
+B_{ij} \, ,
\end{eqnarray}
with $x_i= 2\mu_i/[1-\mu_i-\mu_j+\lambda^{1/2}]$ and the terms
\begin{eqnarray*}
B_{ij} &=& \frac{1-\mu_i-\mu_j} { \lambda^{1/2} } \left[ 4{\rm
Li_{2}}(x_i x_j)- 2{\rm Li_{2}}(-x_i) -2{\rm Li_{2}}(-x_j) +2
\log x_i x_j \log (1-x_ix_j) \right. \nonumber \\
&& \left. \hspace*{2.2cm} - \log x_i \log (1+x_i) - \log x_j \log
(1+x_j) \right] \nonumber \\
&& - 4 \left[ \log (1-x_i x_j)+ \frac{x_i x_j}{1-x_i x_j} \log x_i x_j
\right] \nonumber \\
&& +\frac{ \lambda^{1/2}+\mu_i-\mu_j } {\lambda^{1/2} } \left[
\log (1+x_i) -\frac{x_i}{1+x_i} \log x_i \, \right] \nonumber \\
&& +\frac{\lambda^{1/2}-\mu_i+\mu_j} { \lambda^{1/2} } \left[
\log (1+x_j) -\frac{x_j}{1+x_j} \log x_j \right] \, .
\end{eqnarray*}
The expressions for the SUSY--QCD corrections $\tilde\delta_{ij}^\pm$
are quite involved \cite{h+sqcd}. However, in the limit of large SUSY
parameters $A_{U/D},\mu,M_{\gl}$ and $m_{\tilde U_i/\tilde
D_i}~(i=1,2)$, they approach simple expressions,
{\allowdisplaybreaks
\begin{eqnarray}
\tilde\delta_{UD}^+ & = & -2\delta_U + C_F \frac{\alpha_s}{\pi}\left\{
c_{2U} \frac{m_{\uus_1}^2-m_{\uus_2}^2}{8} \left[ M_{\gl}^2
J(m_{\uus_1}^2,m_{\uus_2}^2,M_{\gl}^2) -
I(m_{\uus_1}^2,m_{\uus_2}^2,M_{\gl}^2) \right] \right. \nonumber \\
& & \left. \hspace*{2.5cm} -c_{2D} \frac{m_{\dds_1}^2-m_{\dds_2}^2}{8} \left[ M_{\gl}^2
J(m_{\dds_1}^2,m_{\dds_2}^2,M_{\gl}^2) -
I(m_{\dds_1}^2,m_{\dds_2}^2,M_{\gl}^2) \right] \right. \nonumber \\
& & \hspace*{2.5cm} - \frac{m_{\dds_1}^2-m_{\uus_1}^2}{8} \left[ M_{\gl}^2
J(m_{\dds_1}^2,m_{\uus_1}^2,M_{\gl}^2) -
I(m_{\dds_1}^2,m_{\uus_1}^2,M_{\gl}^2) \right] \nonumber \\
& & \hspace*{2.5cm} - \frac{m_{\dds_2}^2-m_{\uus_2}^2}{8} \left[ M_{\gl}^2
J(m_{\dds_2}^2,m_{\uus_2}^2,M_{\gl}^2) -
I(m_{\dds_2}^2,m_{\uus_2}^2,M_{\gl}^2) \right] \nonumber \\
& & \hspace*{2.5cm} -M_{\gl} \left(A_U + \frac{\mu}{g_U^A}\right)
\left[s_U^2 c_D^2 I(m_{\uus_1}^2,m_{\dds_1}^2,M_{\gl}^2)
+ c_U^2 c_D^2 I(m_{\uus_2}^2,m_{\dds_1}^2,M_{\gl}^2)
\right.  \nonumber \\
& & \hspace*{5.7cm} + s_U^2 s_D^2 I(m_{\uus_1}^2,m_{\dds_2}^2,M_{\gl}^2)
+ c_U^2 s_D^2 I(m_{\uus_2}^2,m_{\dds_2}^2,M_{\gl}^2) \nonumber \\
& & \hspace*{5.7cm} \left. \left. -
I(m_{\uus_1}^2,m_{\uus_2}^2,M_{\gl}^2) \right] \right\}
\nonumber \\
\tilde\delta_{DU}^+ & = & -2\delta_D + C_F \frac{\alpha_s}{\pi}\left\{
c_{2D} \frac{m_{\dds_1}^2-m_{\dds_2}^2}{8} \left[ M_{\gl}^2
J(m_{\dds_1}^2,m_{\dds_2}^2,M_{\gl}^2) -
I(m_{\dds_1}^2,m_{\dds_2}^2,M_{\gl}^2) \right] \right. \nonumber \\
& & \left. \hspace*{2.5cm} -c_{2U} \frac{m_{\uus_1}^2-m_{\uus_2}^2}{8} \left[ M_{\gl}^2
J(m_{\uus_1}^2,m_{\uus_2}^2,M_{\gl}^2) -
I(m_{\uus_1}^2,m_{\uus_2}^2,M_{\gl}^2) \right] \right. \nonumber \\
& & \hspace*{2.5cm} - \frac{m_{\uus_1}^2-m_{\dds_1}^2}{8} \left[ M_{\gl}^2
J(m_{\dds_1}^2,m_{\uus_1}^2,M_{\gl}^2) -
I(m_{\dds_1}^2,m_{\uus_1}^2,M_{\gl}^2) \right] \nonumber \\
& & \hspace*{2.5cm} - \frac{m_{\uus_2}^2-m_{\dds_2}^2}{8} \left[ M_{\gl}^2
J(m_{\dds_2}^2,m_{\uus_2}^2,M_{\gl}^2) -
I(m_{\dds_2}^2,m_{\uus_2}^2,M_{\gl}^2) \right] \nonumber \\
& & \hspace*{2.5cm} -M_{\gl} \left(A_D + \frac{\mu}{g_D^A}\right)
\left[s_D^2 c_U^2 I(m_{\dds_1}^2,m_{\uus_1}^2,M_{\gl}^2)
+ c_D^2 c_U^2 I(m_{\dds_2}^2,m_{\uus_1}^2,M_{\gl}^2)
\right.  \nonumber \\
& & \hspace*{5.7cm} + s_D^2 s_U^2 I(m_{\dds_1}^2,m_{\uus_2}^2,M_{\gl}^2)
+ c_D^2 s_U^2 I(m_{\dds_2}^2,m_{\uus_2}^2,M_{\gl}^2) \nonumber \\
& & \hspace*{5.7cm} \left. \left. -
I(m_{\dds_1}^2,m_{\dds_2}^2,M_{\gl}^2) \right] \right\}
\nonumber \\
\tilde\delta_{UD}^- & = & -\delta_U-\delta_D \nonumber \\[0.3cm]
& & - \frac{C_F}{2} \frac{\alpha_s}{\pi} M_{\gl} \left\{
\left(A_D + \frac{\mu}{g_D^A}\right) \left[
  c_U^2 s_D^2 I(m_{\dds_1}^2,m_{\uus_1}^2,M_{\gl}^2)
+ s_U^2 s_D^2 I(m_{\dds_1}^2,m_{\uus_2}^2,M_{\gl}^2) \right.\right. \nonumber \\
& & \hspace*{4.7cm} + c_U^2 c_D^2 I(m_{\dds_2}^2,m_{\uus_1}^2,M_{\gl}^2)
+ s_U^2 c_D^2 I(m_{\dds_2}^2,m_{\uus_2}^2,M_{\gl}^2) \nonumber \\
& & \left. \hspace*{4.7cm}
-I(m_{\dds_1}^2,m_{\dds_2}^2,M_{\gl}^2)\right] \nonumber \\
& & \hspace*{2.1cm} + \left(A_U + \frac{\mu}{g_U^A}\right) \left[
  s_U^2 c_D^2 I(m_{\dds_1}^2,m_{\uus_1}^2,M_{\gl}^2)
+ c_U^2 c_D^2 I(m_{\dds_1}^2,m_{\uus_2}^2,M_{\gl}^2) \right. \nonumber \\
& & \hspace*{4.7cm}
+ s_U^2 s_D^2 I(m_{\dds_2}^2,m_{\uus_1}^2,M_{\gl}^2)
+ c_U^2 s_D^2 I(m_{\dds_2}^2,m_{\uus_2}^2,M_{\gl}^2) \nonumber \\
& & \left. \left. \hspace*{4.7cm}
-I(m_{\uus_1}^2,m_{\uus_2}^2,M_{\gl}^2)\right] \right\} \, ,
\label{eq:sqcdinfty}
\end{eqnarray}}
with the function $I$ of Eq.~(\ref{eq:ii}) and
\begin{equation}
J(a,b,c) = \frac{\partial I(a,b,c)}{\partial c} = \frac{1}{a-b} \left\{
\frac{\displaystyle a\log\frac{a}{c}}{(a-c)^2} - \frac{\displaystyle
b\log\frac{b}{c}}{(b-c)^2}\right\} + \frac{1}{(a-c)(b-c)} \, .
\label{eq:jj}
\end{equation}
The terms $\delta_Q ~ (Q=U,D)$ are related to the $\Delta_{U/D}$
terms, respectively,
\begin{eqnarray}
\delta_Q = \Delta_Q \left[ 1+\frac{1}{\left(g_Q^A\right)^2}\right]
\qquad (Q=U,D) \, .
\end{eqnarray}
For the third generation, $\Delta_U$ is given by $\Delta_t$ of
Eq.~(\ref{eq:deltat}) and $\Delta_D$ by $\Delta_b^{QCD}$ of
Eq.~(\ref{eq:deltab}).

The use of the effective up- and down-type Yukawa couplings $\tilde g_U^A,
\tilde g_D^A$ at LO leads to the subtraction of the $\delta_U$ and $\delta_D$
terms in Eq.~(\ref{eq:sqcdinfty}), leaving the SUSY-remainders
\begin{eqnarray}
\tilde \delta_{UD}^+ & \to & \tilde \delta_{UD,rem}^+ = \tilde \delta_{UD}^+
+ 2\delta_U \nonumber \\
\tilde \delta_{DU}^+ & \to & \tilde \delta_{DU,rem}^+ = \tilde \delta_{DU}^+
+ 2\delta_D \nonumber \\
\tilde \delta_{UD}^- & \to & \tilde \delta_{UD,rem}^- = \tilde \delta_{UD}^-
+ \delta_U + \delta_D
\end{eqnarray}
in the SUSY--QCD corrections to the charged Higgs decay of
Eq.~(\ref{eq:h+udsqcd0}). This yields the improved expression of the
partial decay width
\begin{eqnarray}
\Gamma [H^+\rightarrow U\bar{D}\,] &=& \frac{3 G_F
M_{H^\pm}}{4\sqrt{2}\pi}
|V_{UD}|^2 \, \lambda^{1/2} \, \Big\{ \nonumber \\
& & \left. (1-\mu_U -\mu_D) \left[
M_U^2 \tilde g_U^A
\left( 1+ \frac{4}{3} \frac{\alpha_s}{\pi} \delta_{UD}^+ \right)
\Big( \tilde g_U^A + g_U^A \tilde\delta_{UD,rem}^+ \Big) \right. \right.
\nonumber \\
& & \left. \left. \qquad\qquad\qquad\quad
+ M_D^2 \tilde g_D^A \left( 1+ \frac{4}{3}
\frac{\alpha_s} {\pi} \delta_{DU}^+ \right)
\Big( \tilde g_D^A + g_D^A \tilde\delta_{DU,rem}^+ \Big)
\right]
\right. \label{eq:h+udsqcd} \\
& & \left. -4M_UM_D \sqrt{\mu_U \mu_D}
\left( 1+ \frac{4}{3}
\frac{\alpha_s}{\pi} \delta_{UD}^- \right)
\left[ \tilde g_U^A \tilde g_D^A + \frac{\tilde g_U^A g_D^A + g_U^A
\tilde g_D^A}{2} \tilde\delta_{UD,rem}^- \right]
\right\} \nonumber
\end{eqnarray}
for all charged Higgs decays into quarks. We extended our NNLO
calculation of the $\Delta_Q$ $(Q=t,b)$ contributions to the charm
quarks and included $\Delta_s$ terms in the effective strange Yukawa
coupling \cite{deltabnnlo2}. We have implemented the effective
Yukawa-coupling factors $\tilde g_U^A, \tilde g_D^A$ in every term
emerging from the LO amplitude during the contraction with the NLO
amplitude, in complete analogy to the computation of neutral Higgs-boson
decays into bottom quarks, where this prescription avoids artificial
singularities \cite{deltab1}.

For large charged Higgs masses, the above expressions for the partial
widths simplify. Due to the chiral structure of the charged Higgs
vertex, the QCD corrections approach the QCD corrections to the scalar
current correlator, which are known up to N$^4$LO \cite{hqcd4}. In the
large Higgs mass regime $M_{H^\pm} \gg M_U + M_D$, the improved partial
decay width is given by \cite{hud1,hud,review}
\begin{eqnarray}
\Gamma [\,H^{+} \to \, U{\overline{D}}\,] & = &
\frac{3 G_F M_{H^\pm}}{4\sqrt{2}\pi} \, \left| V_{UD} \right|^2 \,
\left[ \overline{m}_U^2 \tilde g_U^{A} \left( \tilde g_U^{A}
+ g_U^{A} \tilde\delta_{UD,rem}^+\right) \right.
\label{eq:hcud}
\\
& & \left. \hspace*{3cm} + \overline{m}_D^2 \tilde g_D^{A}
\left( \tilde g_D^{A}
+ g_D^{A} \tilde\delta_{DU,rem}^+\right) \right] (1 + \delta_{\rm QCD})
\, , \nonumber
\end{eqnarray}
where $\overline{m}_{U,D}$ denote the $\overline{\rm MS}$ masses
evaluated at the scale of the charged Higgs-mass $M_{H^\pm}$.  In
this expression, we have used the effective Yukawa coupling $\tilde
g_{U/D}^A$ in every contribution that emerges from the LO amplitude, as
in Eq.~(\ref{eq:h+udsqcd}), but we use the LO Yukawa-coupling factors for
the SUSY-remainder itself at the amplitude level. The QCD corrections
$\delta_{\rm QCD}$ are given by \cite{hqcd4}
\begin{eqnarray}
\delta_{\rm QCD} & = & 5.67 \frac{\alpha_s (M_{H^\pm})}{\pi} + (35.94 -
1.36 N_F) \left[ \frac{\alpha_s (M_{H^\pm})}{\pi} \right]^2 \nonumber \\
& & + (164.14 - 25.77 N_F + 0.259 N_F^2) \left[
\frac{\alpha_s(M_{H^\pm})}{\pi} \right]^3 \nonumber \\
& & +(39.34-220.9 N_F+9.685 N_F^2-0.0205 N_F^3) \left[
\frac{\alpha_s(M_{H^\pm})}{\pi} \right]^4 \, ,
\end{eqnarray}
where large logarithms are absorbed in the $\overline{\rm MS}$ masses
$\overline{m}_{U,D}$ at the charged-Higgs mass scale and we have used
$N_F=5$.

\begin{figure}
\begin{center}
\vspace*{-1.0cm}
\epsfig{file=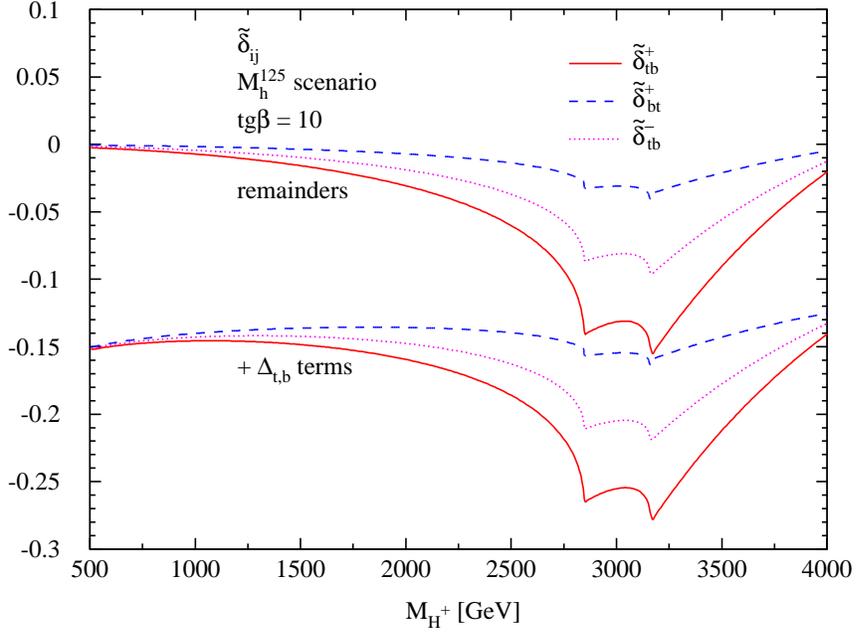,%
        bbllx=30pt,bblly=350pt,bburx=520pt,bbury=650pt,%
        scale=0.6}
\vspace*{2.5cm}

\epsfig{file=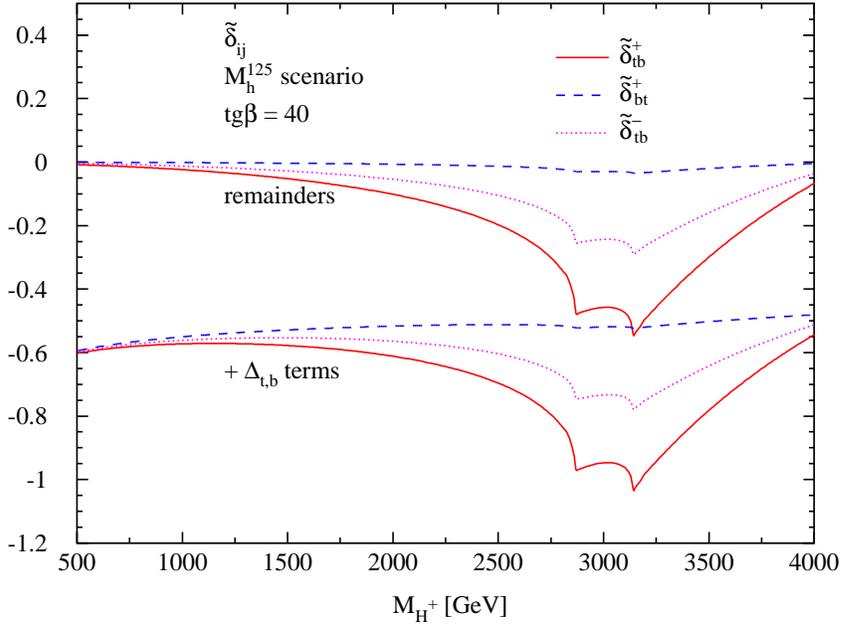,%
        bbllx=30pt,bblly=350pt,bburx=520pt,bbury=650pt,%
        scale=0.6}
\end{center}
\vspace*{2.0cm}
\caption{\it Individual NLO coefficients
$\tilde\delta_{ij}^\pm~(i,j=t,b)$ of the decay width $\Gamma(H^+\to
t\bar b)$ as a function of the charged Higgs mass with and without the
$\Delta_{t/b}$ contributions in the $M_h^{125}$ scenario. The upper
curves show the sizes of the SUSY-remainders, while the lower curves
correspond to the full corrections including the $\Delta_{t/b}$
contributions, without using the effective Yukawa couplings.}
\label{fg:susyrem}
\end{figure}
We have implemented these improved results for the partial charged Higgs
decay widths into heavy quarks, $H^+ \to t\bar b, c\bar b, c\bar s$,
using the corresponding expressions for $\Delta_t$, $\Delta_b$, $\Delta_c$
and $\Delta_s$ up to NNLO in the code {\tt Hdecay} \cite{hdecay}. This
allows us to predict these partial charged Higgs decay widths with
approximate NNLO SUSY--QCD precision within the MSSM. In our numerical
analysis, for the SUSY-remainders $\tilde\delta_{ij}^\pm$ themselves we
have consistently used the top pole mass and the derived bottom mass, in
accordance with their treatment in the stop and sbottom sectors
\cite{gganlo}. Fig.~\ref{fg:susyrem} shows the size of the individual
NLO SUSY coefficients $\tilde\delta_{ij}^\pm$ $(i,j=t,b)$ as a function
of the charged Higgs mass within the $M_h^{125}$ benchmark scenario for
two values of $\tgb=10, 40$. The upper curves correspond to the
SUSY-remainders $\tilde\delta^\pm_{ij,rem}$ after absorbing the dominant
$\Delta_{t/b}$ terms in the effective Yukawa coupling factors $\tilde
g_{t/b}^A$, while the lower curves exhibit the full NLO SUSY--QCD
corrections without using the effective Yukawa couplings, i.e.~the
$\Delta_{b,t}$ terms are part of the fixed-order NLO correction. It is
clearly visible that for charged Higgs masses up to about 2 TeV the
SUSY-remainders are in the percent range and below so that the effective
Yukawa-coupling factors $\tilde g_{t,b}$ alone yield a reliable
approximation of the full NLO results. It should be noted that also the
$\Delta_t$ terms contained in the effective top Yukawa coupling factor
$\tilde g_t^A$ provide an excellent approximation of the NLO SUSY--QCD
corrections emerging from the top-Yukawa coupling. This can be
understood from the leading terms in the large SUSY-mass expansion of
Eq.~(\ref{eq:sqcdinfty}): The remainder terms are either not enhanced
for large values of $\tgb$ or the $\tgb$-enhanced contributions are
proportional to differences of the function $I$ of Eq.~(\ref{eq:ii})
that are small for nearly degenerate soft SUSY-breaking squark-mass
parameters. Only for larger Higgs masses close to and above the virtual
stop and sbottom thresholds the remainders turn out to be more sizable,
as expected for kinematical reasons. For a charged Higgs mass below about
2 TeV, this result can also be used for an NNLO approximation of the
SUSY--QCD corrections, since the SUSY remainder of the NNLO corrections
is expected to be even further suppressed in the limit of large
SUSY-masses.

\begin{figure}[htbp]
\begin{center}
\vspace*{-1.0cm}
\epsfig{file=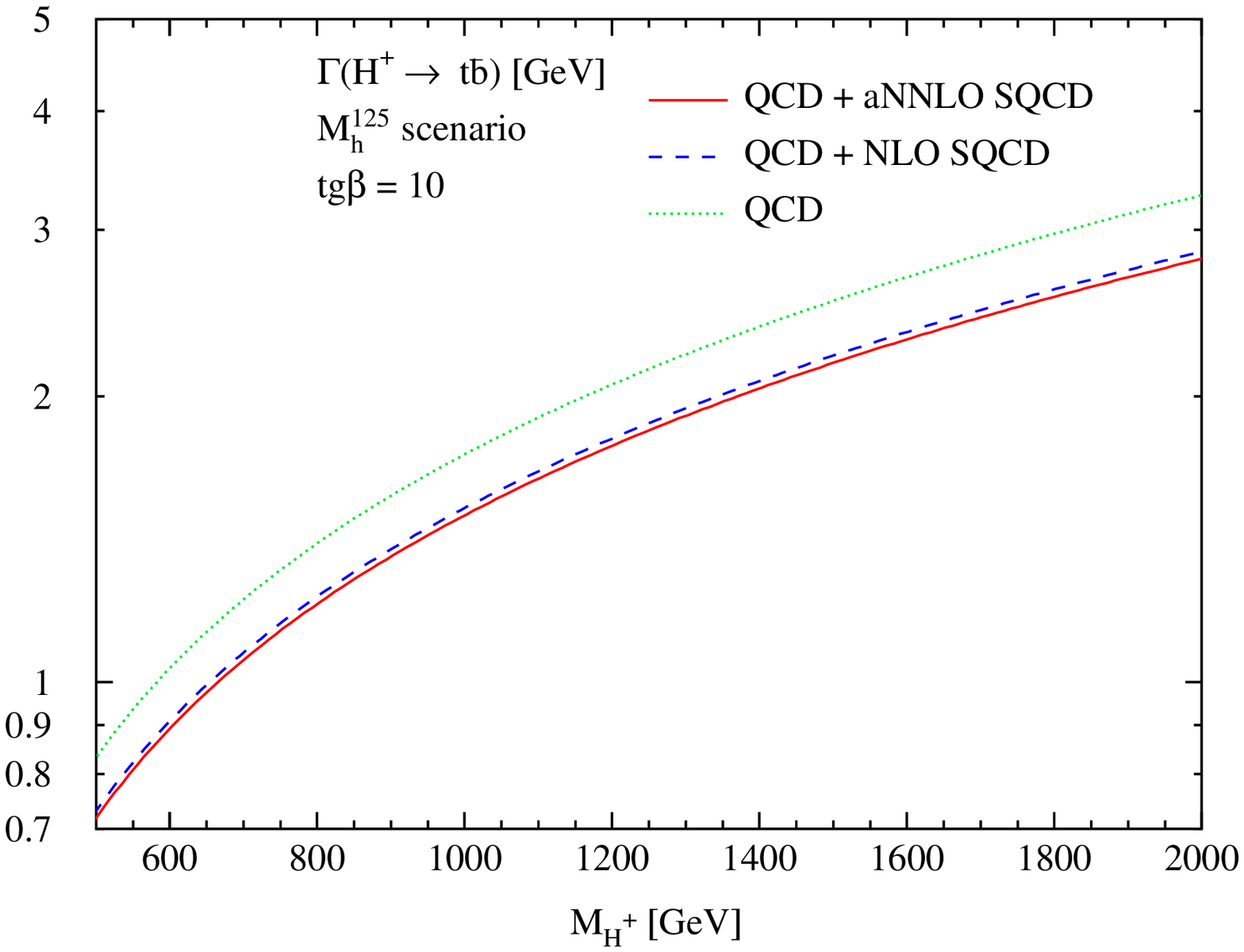,%
        bbllx=30pt,bblly=350pt,bburx=520pt,bbury=650pt,%
        scale=0.6}
\vspace*{2.5cm}

\epsfig{file=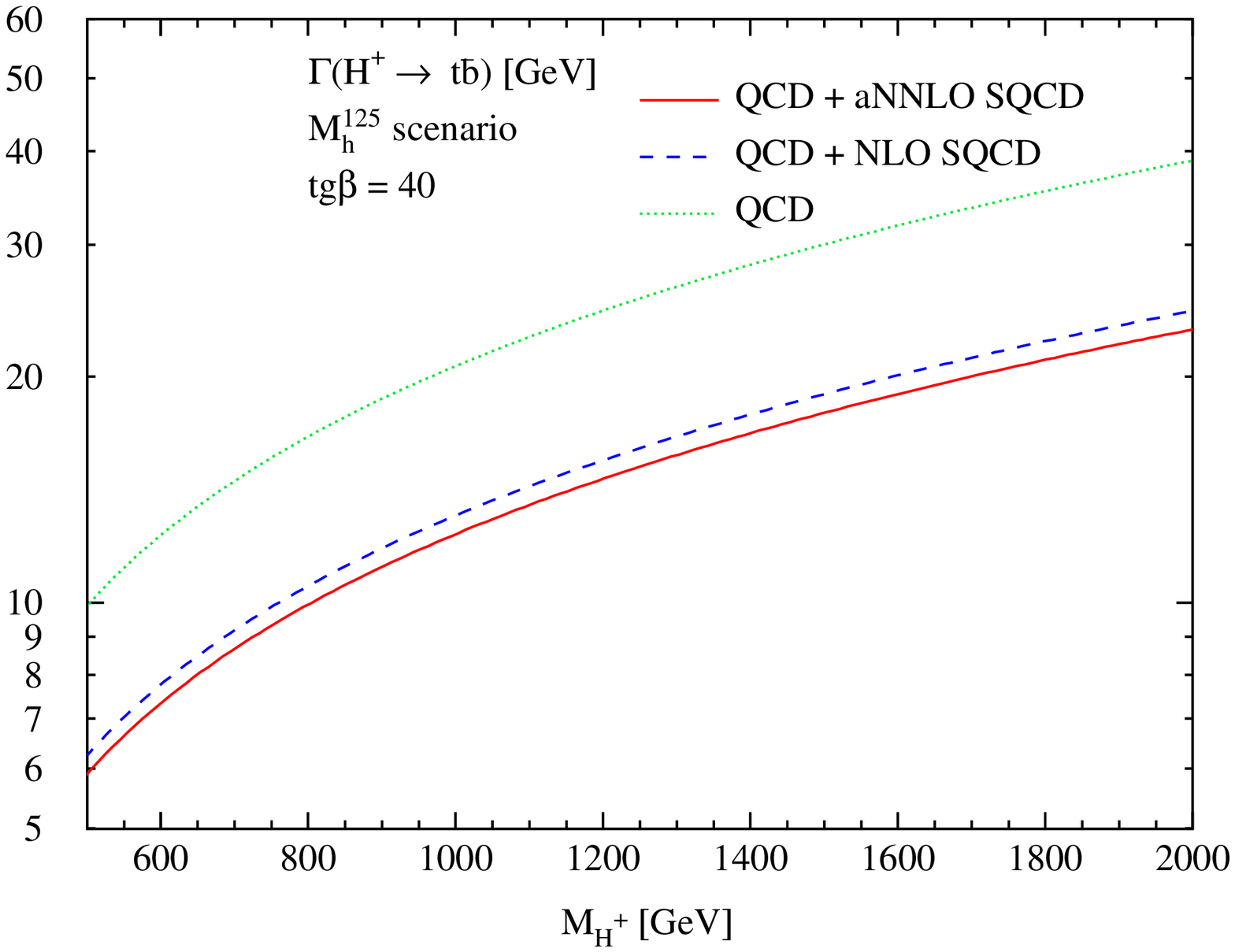,%
        bbllx=30pt,bblly=350pt,bburx=520pt,bbury=650pt,%
        scale=0.6}
\end{center}
\vspace*{2.0cm}
\caption{\it Partial decay width of $H^+\to t\bar b$ at different orders
of the SUSY--QCD corrections. The dotted (green) lines include the usual
QCD corrections involving the $\overline{\rm MS}$ Yukawa couplings, but
no genuine SUSY--QCD corrections. The dashed (blue) lines involve the
full NLO SUSY--QCD remainders in addition to the effective Yukawa
couplings that have been used at the 1-loop level in SUSY--QCD. The full
(red) lines display the approximate results (aNNLO) including the 2-loop
SUSY--QCD corrections to the effective Yukawa couplings.}
\label{fg:gamma_tb}
\end{figure}

\begin{figure}[htbp]
\begin{center}
\vspace*{-1.0cm}
\epsfig{file=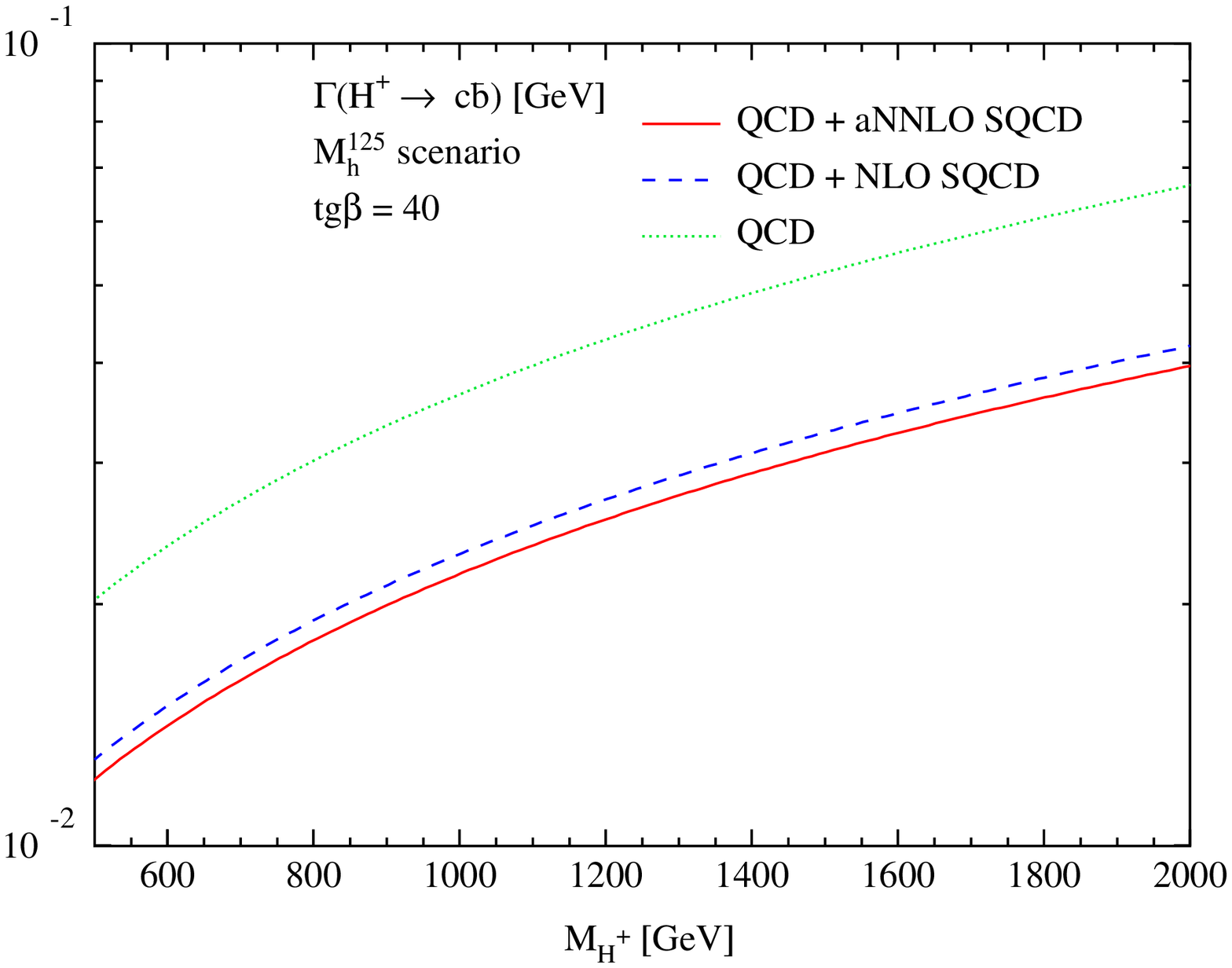,%
        bbllx=30pt,bblly=350pt,bburx=520pt,bbury=650pt,%
        scale=0.6}
\vspace*{2.5cm}

\epsfig{file=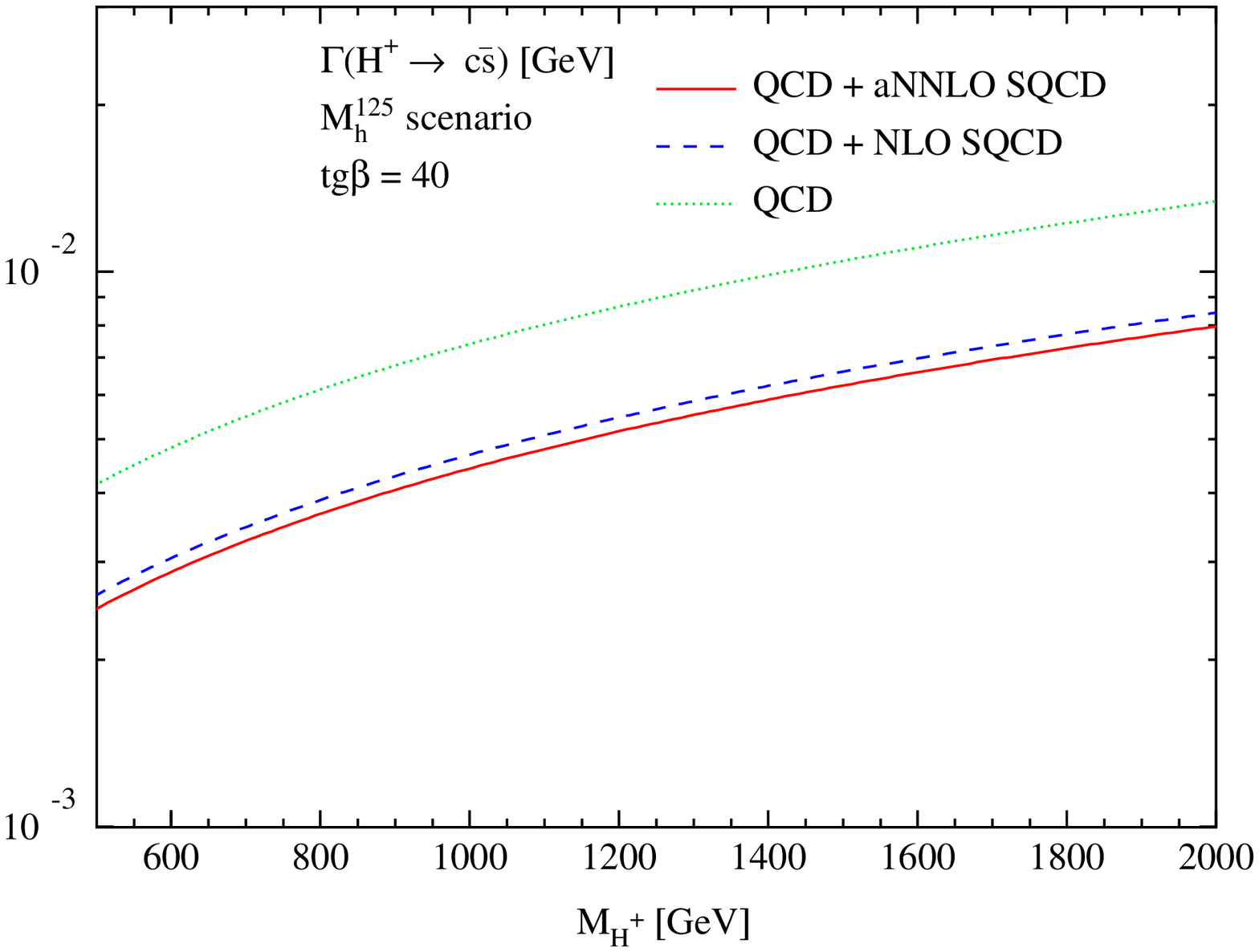,%
        bbllx=30pt,bblly=350pt,bburx=520pt,bbury=650pt,%
        scale=0.6}
\end{center}
\vspace*{2.0cm}
\caption{\it Same as Fig.~\ref{fg:gamma_tb} but for $H^+\to c\bar b$
(top) and $H^+\to c\bar s$ (bottom) for $\tgb=40$.}
\label{fg:gamma_cx}
\end{figure}

In Fig.~\ref{fg:gamma_tb} we show the partial decay width $\Gamma(H^+\to
t\bar b)$ with pure QCD corrections and NLO/approximate NNLO (in terms
of the effective Yukawa couplings) SUSY--QCD corrections as a function
of the charged Higgs mass. We show the results for two values of
$\tgb=10, 40$ and for the central scale choices. While the NLO SUSY--QCD
corrections modify the partial decay width substantially, the
approximate NNLO corrections are still relevant. They are at the level
of a few percent over the full charged Higgs-mass range and reduce the
theoretical uncertainties related to the scale choice to the percent
level.
The analogous picture emerges for the subleading decay modes $H^+\to
c\bar b$ and $H^+\to c\bar s$, as can be inferred from
Fig.~\ref{fg:gamma_cx} that includes the analogous SUSY--QCD corrections
to the charm and strange Yukawa couplings.

\section{Conclusions} \label{sc:conclusions}
In this work we have re-examined the charged Higgs-boson decays into
heavy quarks within the MSSM. We rederived the genuine NLO SUSY--QCD
corrections and combined them with the usual QCD corrections in a
factorized form. We discussed the role of introducing effective
down-type and up-type Yukawa couplings that constitute the dominant
contributions to the genuine SUSY--QCD corrections and analyzed the
accuracy of this approximation to the full contributions. Since the
SUSY-remainder beyond the effective Yukawa couplings turned out
to be small at NLO and we expect the same to be valid beyond NLO, we can
obtain approximate NNLO results for the genuine SUSY--QCD corrections by
evaluating these corrections to the effective Yukawa couplings. This
amounts to calculating the down-type $\Delta_D$ and up-type $\Delta_U$
corrections at NNLO. The NNLO calculation to $\Delta_U$ is new in our
work. Combining these approximate NNLO SUSY--QCD corrections with the
usual QCD corrections that are known up to N$^4$LO for large Higgs
masses, we have obtained an approximate NNLO result for charged
Higgs-boson decays into heavy quarks. We have applied our calculation to
the charged Higgs-boson decays $H^+\to t\bar b, c\bar b, c\bar s$ in
particular, but have implemented these corrections in the public code
{\tt Hdecay} \cite{hdecay} for all charged Higgs decays into quarks.
This makes the theoretical predictions of the charged Higgs decays more
precise than previously. The new implementations in {\tt Hdecay} will be
made public soon.

Our calculation can easily be extended to non-minimal supersymmetric
models as e.g.~the next-to-MSSM (NMSSM). This requires taking into
account an extended Higgs sector involving more mixing angles than in the
MSSM. The translation is straightforward, as demonstrated for the
public code {\tt NMSSMCALC} \cite{nmssmcalc} that computes the
radiatively corrected Higgs masses and decay width including all
available higher-order corrections. \\

\noindent
{\bf Acknowledgements} \\
F.K.~is funded by the Deutsche Forschungsgemeinschaft (DFG, German
Research Foundation) under Germany's Excellence Strategy -- EXC-2123
QuantumFrontiers -- 390837967. The work of M.M.~is supported by the DFG
Collaborative Research Center TRR257 ``Particle Physics Phenomenology
after the Higgs Discovery''.


\begin{thebibliography}{99}

\bibitem{pdg}
P.A.~Zyla \textit{et al.} [Particle Data Group],
PTEP \textbf{2020} (2020) no.8, 083C01.

\bibitem{higgs}
  P.~W.~Higgs,
  Phys.\ Lett.\  {\bf 12} (1964) 132,
  Phys.\ Rev.\ Lett.\  {\bf 13} (1964) 508
and
  Phys.\ Rev.\  {\bf 145} (1966) 1156;
  F.~Englert and R.~Brout,
  Phys.\ Rev.\ Lett.\  {\bf 13} (1964) 321;
  G.~S.~Guralnik, C.~R.~Hagen and T.~W.~Kibble,
  Phys.\ Rev.\ Lett.\  {\bf 13} (1964) 585;
  T.~W.~B.~Kibble,
  Phys.\ Rev.\  {\bf 155} (1967) 1554.

\bibitem{discovery}
  G.~Aad {\it et al.}  [ATLAS Coll.],
  Phys.\ Lett.\ {\bf B716} (2012) 1;
  S.~Chatrchyan {\it et al.}  [CMS Coll.],
  Phys.\ Lett.\ {\bf B716} (2012) 30.

\bibitem{couplings}
  G.~Aad {\it et al.} [ATLAS and CMS Collaborations],
  JHEP {\bf 1608} (2016) 045;
 G.~Aad {\it et al.} [ATLAS Collaboration],
ATLAS-CONF-2019-005;
A.M.~Sirunyan \textit{et al.} [CMS Collaboration],
JHEP \textbf{01} (2021) 148.

\bibitem{sm}
S.L.~Glashow, Nucl.~Phys.~{\bf 22} (1961) 579;
S.~Weinberg, Phys.~Rev.~Lett.~{\bf 19} (1967) 1264;
A.~Salam, Conf.~Proc.~{\bf C680519} (1968) 367.

\bibitem{smren}
  G.~'t Hooft,
  Nucl.\ Phys.\ {\bf B35} (1971) 167;
  G.~'t Hooft and M.~J.~G.~Veltman,
  Nucl.\ Phys.\ {\bf B44} (1972) 189.

\bibitem{unitarity}
  C.~H.~Llewellyn Smith,
  Phys.\ Lett.\  {\bf 46B} (1973) 233;
  J.~M.~Cornwall, D.~N.~Levin and G.~Tiktopoulos,
  Phys.\ Rev.\ {\bf D10} (1974) 1145,
   Erratum: [Phys.\ Rev.\ {\bf D11} (1975) 972];
  B.~W.~Lee, C.~Quigg and H.~B.~Thacker,
  Phys.\ Rev.\ Lett.\  {\bf 38} (1977) 883
and
  Phys.\ Rev.\ {\bf D16} (1977) 1519.

\bibitem{hmassbounds}
  N.~Cabibbo, L.~Maiani, G.~Parisi and R.~Petronzio,
  Nucl.\ Phys.\ {\bf B158} (1979) 295;
  M.~S.~Chanowitz, M.~A.~Furman and I.~Hinchliffe,
  Phys.\ Lett.\  {\bf 78B} (1978) 285;
  R.~A.~Flores and M.~Sher,
  Phys.\ Rev.\ {\bf D27} (1983) 1679;
  M.~Lindner,
  Z.\ Phys.\ {\bf C31} (1986) 295;
  A.~Hasenfratz, K.~Jansen, C.~B.~Lang, T.~Neuhaus and H.~Yoneyama,
  Phys.\ Lett.\ {\bf B199} (1987) 531;
  J.~Kuti, L.~Lin and Y.~Shen,
  Phys.\ Rev.\ Lett.\  {\bf 61} (1988) 678;
  M.~L\"uscher and P.~Weisz,
  Nucl.\ Phys.\ {\bf B318} (1989) 705;
  M.~Sher,
  Phys.\ Rept.\ {\bf 179} (1989) 273
and
  Phys.\ Lett.\ {\bf B317} (1993) 159,
   Addendum: [Phys.\ Lett.\ {\bf B331} (1994) 448];
  G.~Altarelli and G.~Isidori,
  Phys.\ Lett.\ {\bf B337} (1994) 141;
  J.~A.~Casas, J.~R.~Espinosa and M.~Quiros,
  Phys.\ Lett.\ {\bf B342} (1995) 171;
  G.~Degrassi, S.~Di Vita, J.~Elias-Miro, J.~R.~Espinosa, G.~F.~Giudice,
  G.~Isidori and A.~Strumia,
  JHEP {\bf 1208} (2012) 098.

\bibitem{metastability}
  J.~R.~Espinosa and M.~Quiros,
  Phys.\ Lett.\ {\bf B353} (1995) 257;
J.~Elias-Miro, J.~R.~Espinosa, G.~F.~Giudice, G.~Isidori, A.~Riotto and A.~Strumia,
Phys. Lett. \textbf{B709} (2012) 222;
  D.~Buttazzo, G.~Degrassi, P.~P.~Giardino, G.~F.~Giudice, F.~Sala,
  A.~Salvio and A.~Strumia,
  JHEP {\bf 1312} (2013) 089;
A.~V.~Bednyakov, B.~A.~Kniehl, A.~F.~Pikelner and O.~L.~Veretin,
Phys. Rev. Lett. \textbf{115} (2015) no.20, 201802.

\bibitem{hierarchy}
  E.~Gildener and S.~Weinberg,
  Phys.\ Rev.\ {\bf D13} (1976) 3333;
  S.~Weinberg,
  Phys.\ Rev.\ {\bf D13} (1976) 974
and
  Phys.\ Rev.\ {\bf D19} (1979) 1277;
  L.~Susskind,
  Phys.\ Rev.\ {\bf D20} (1979) 2619.

\bibitem{susy}
  Y.~A.~Golfand and E.~P.~Likhtman,
  JETP Lett.\  {\bf 13} (1971) 323
   [Pisma Zh.\ Eksp.\ Teor.\ Fiz.\  {\bf 13} (1971) 452];
  D.~V.~Volkov and V.~P.~Akulov,
  Phys.\ Lett.\  {\bf 46B} (1973) 109;
  J.~Wess and B.~Zumino,
  Nucl.\ Phys.\ {\bf B70} (1974) 39.

\bibitem{sw2sgut}
  S.~Dimopoulos, S.~Raby and F.~Wilczek,
  Phys.\ Rev.\ {\bf D24} (1981) 1681;
  L.~E.~Ibanez and G.~G.~Ross,
  Phys.\ Lett.\  {\bf 105B} (1981) 439.

\bibitem{mssm}
 P.~Fayet,
  Nucl.\ Phys.\ B {\bf 90} (1975) 104,
  Phys.\ Lett.\ B {\bf 64} (1976) 159
and
  Phys.\ Lett.\ B {\bf 69} (1977) 489.

\bibitem{mssm1} For reviews on supersymmetric theories, see
  P.~Fayet and S.~Ferrara,
  Phys.\ Rept.\  {\bf 32} (1977) 249;
  H.~P.~Nilles,
  Phys.\ Rept.\  {\bf 110} (1984) 1;
  R.~Barbieri,
  Riv.\ Nuovo Cim.\  {\bf 11N4} (1988) 1;
  H.~E.~Haber and G.~L.~Kane,
  Phys.\ Rept.\  {\bf 117} (1985) 75.

\bibitem{benchmark}
  E.~Bagnaschi {\it et al.},
  Eur.~Phys.~J.~{\bf C79} (2019) no.7,  617.

\bibitem{gganlo}
E.~Bagnaschi, L.~Fritz, S.~Liebler, M.~M\"uhlleitner, T.T.D.~Nguyen
and M.~Spira,
JHEP \textbf{03} (2023) 124.

\bibitem{rgi}
M.~Carena, H.E.~Haber, S.~Heinemeyer, W.~Hollik, C.E. Wagner and
G.~Weiglein,
Nucl.~Phys.~{\bf B580} (2000) 29.

\bibitem{sbottom}
A.~Brignole, G.~Degrassi, P.~Slavich and F.~Zwirner,
Nucl.\ Phys.\ \textbf{B643} (2002) 79;
S.~Heinemeyer, W.~Hollik, H.~Rzehak and G.~Weiglein,
Eur.\ Phys.\ J.\ \textbf{C39} (2005) 465;
S.~Heinemeyer, H.~Rzehak and C.~Schappacher,
Phys.\ Rev.\ \textbf{D82} (2010), 075010;
G.~Degrassi and P.~Slavich,
JHEP \textbf{11} (2010) 044.

\bibitem{hdecay}
  A.~Djouadi, J.~Kalinowski and M.~Spira,
  Comput.\ Phys.\ Commun.\ {\bf 108} (1998) 56;
  A.~Djouadi, M.~M.~M\"uhlleitner and M.~Spira,
  Acta Phys.\ Polon.\ {\bf B38} (2007) 635;
  A.~Djouadi, J.~Kalinowski, M.~M\"uhlleitner and M.~Spira,
  Comput.\ Phys.\ Commun.\  {\bf 238} (2019) 214.

\bibitem{1OFF}
  A.~Djouadi, J.~Kalinowski and P.~M.~Zerwas,
  Z.\ Phys.\ {\bf C70} (1996) 435;
  S.~Moretti and W.~J.~Stirling,
  Phys.\ Lett.\ {\bf B347} (1995) 291,
   Erratum: [Phys.\ Lett.\ {\bf B366} (1996) 451].

\bibitem{deltab}
  L.J.~Hall, R.~Rattazzi and U.~Sarid,
  Phys.\ Rev.\ {\bf D50}, 7048 (1994);
  R.~Hempfling,
  Phys.\ Rev.\ {\bf D49}, 6168 (1994);
  M.~Carena, M.~Olechowski, S.~Pokorski and C.E.M.~Wagner,
  Nucl.\ Phys.\ {\bf B426} (1994) 269;
  D.M.~Pierce, J.A.~Bagger, K.T.~Matchev and R.-J.~Zhang,
  Nucl.\ Phys.\ {\bf B491}, 3 (1997);
  J.~Guasch, W.~Hollik and S.~Pe\~naranda,
  Phys.\ Lett.\ {\bf B515} (2001) 367;
  G.~D'Ambrosio, G.F.~Giudice, G.~Isidori and A.~Strumia,
  Nucl.\ Phys.\ {\bf B645}, 155 (2002);
  A.~J.~Buras, P.H.~Chankowski, J.~Rosiek and L.~Slawianowska,
  Nucl.\ Phys.\ {\bf B659}, 3 (2003);
  V.~Barger, H.E.~Logan and G.~Shaughnessy,
  Phys.\ Rev.\ {\bf D79}, 115018 (2009);
  N.D.~Christensen, T.~Han and S.~Su,
  Phys.\ Rev.\ {\bf D85}, 115018 (2012);
  M.S.~Carena, D.~Garcia, U.~Nierste and C.E.M.~Wagner,
  Nucl.\ Phys.\ {\bf B577}, 88 (2000).

\bibitem{deltab1}
  J.~Guasch, P.~H\"afliger and M.~Spira,
  Phys.\ Rev.\ {\bf D68}, 115001 (2003).

\bibitem{deltabnnlo}
  D.~Noth and M.~Spira,
  Phys.\ Rev.\ Lett.\  {\bf 101}, 181801 (2008)
and
  JHEP {\bf 1106}, 084 (2011).

\bibitem{deltabnnlo1}
  L.~Mihaila and C.~Reisser,
  JHEP {\bf 1008}, 021 (2010);
  A.~Crivellin and C.~Greub,
  Phys.\ Rev.\ {\bf D87} (2013) 015013,
   Erratum: [Phys.\ Rev.\ {\bf D87} (2013) 079901].

\bibitem{deltabnnlo2}
M.~Ghezzi, S.~Glaus, D.~M\"uller, T.~Schmidt and M.~Spira,
Eur.\ Phys.\ J.\ \textbf{C81} (2021) no.3, 259.

\bibitem{mihailazerf}
L.~Mihaila and N.~Zerf,
JHEP \textbf{05} (2017) 019.

\bibitem{let}
J.R.~Ellis, M.K.~Gaillard and D.V.~Nanopoulos,
Nucl.\ Phys.\  {\bf B106} (1976) 292;
M.A.~Shifman, A.I.~Vainshtein, M.B.~Voloshin and V.I.~Zakharov,
Sov.\ J.\ Nucl.\ Phys.\  {\bf 30} (1979) 711
[Yad.\ Fiz.\  {\bf 30} (1979) 1368];
M.\,Spira, A.\,Djouadi, D.\,Graudenz and P.M.\,Zerwas,
Nucl.\, Phys.\, {\bf B453} (1995) 17;
B.A.\,Kniehl and M.\,Spira, Z.\,Phys.\,{\bf C69} (1995) 77;
W.\,Kilian,
Z.\,Phys.\,{\bf C69} (1995) 89.

\bibitem{hud1}
  A.~Mendez and A.~Pomarol,
  Phys.\ Lett.\ {\bf B252} (1990) 461;
  C.~S.~Li and R.~J.~Oakes,
  Phys.\ Rev.\ {\bf D43} (1991) 855.

\bibitem{hud}
  A.~Djouadi and P.~Gambino,
  Phys.\ Rev.\ {\bf D51} (1995) 218,
   Erratum: [Phys.\ Rev.\ {\bf D53} (1996) 4111].

\bibitem{h+sqcd}
  A.~Dabelstein,
  Nucl.\ Phys.\ {\bf B456} (1995) 25;
  R.~A.~Jimenez and J.~Sola,
  Phys.\ Lett.\ {\bf B389} (1996) 53;
  J.~A.~Coarasa Perez, R.~A.~Jimenez and J.~Sola,
  Phys.\ Lett.\ {\bf B389} (1996) 312.

\bibitem{hqcd4}
  E.~Braaten and J.~P.~Leveille,
  Phys.\ Rev.\ {\bf D22} (1980) 715;
  N.~Sakai,
  Phys.\ Rev.\ {\bf D22} (1980) 2220;
  T.~Inami and T.~Kubota,
  Nucl.\ Phys.\ {\bf B179} (1981) 171;
  M.~Drees and K.~I.~Hikasa,
  Phys.\ Lett.\ {\bf B240} (1990) 455,
   Erratum: [Phys.\ Lett.\ {\bf B262} (1991) 497]
and
  Phys.\ Rev.\ {\bf D41} (1990) 1547;
  S.~G.~Gorishnii, A.~L.~Kataev and S.~A.~Larin,
  Sov.\ J.\ Nucl.\ Phys.\  {\bf 40} (1984) 329
   [Yad.\ Fiz.\  {\bf 40} (1984) 517];
  S.~G.~Gorishnii, A.~L.~Kataev, S.~A.~Larin and L.~R.~Surguladze,
  Mod.\ Phys.\ Lett.\ {\bf A5} (1990) 2703
and
  Phys.\ Rev.\ {\bf D43} (1991) 1633;
  A.~L.~Kataev and V.~T.~Kim,
  Mod.\ Phys.\ Lett.\ {\bf A9} (1994) 1309;
  L.~R.~Surguladze,
  Phys.\ Lett.\ {\bf B341} (1994) 60;
  K.~G.~Chetyrkin,
  Phys.\ Lett.\ {\bf B390} (1997) 309;
P.~A.~Baikov, K.~G.~Chetyrkin and J.~H.~K\"uhn,
Phys. Rev. Lett. \textbf{96} (2006) 012003;
F.~Herzog, B.~Ruijl, T.~Ueda, J.~A.~M.~Vermaseren and A.~Vogt,
JHEP \textbf{08} (2017), 113.

\bibitem{review}
  M.~Spira,
  Fortsch.\ Phys.\  {\bf 46} (1998) 203
and
  Prog.\ Part.\ Nucl.\ Phys.\  {\bf 95} (2017) 98;
  A.~Djouadi,
  Phys.\ Rept.\  {\bf 459} (2008) 1.

\bibitem{nmssmcalc}
J.~Baglio, R.~Gr\"ober, M.~M\"uhlleitner, D.~T.~Nhung, H.~Rzehak,
M.~Spira, J.~Streicher and K.~Walz,
Comput. Phys. Commun. \textbf{185} (2014) no.12, 3372.

\end{thebibliography}
\end{document}